\documentclass[twocolumn,amsmath,amssymb,aps,prb,longbibliography]{revtex4-1}
\usepackage{graphicx}
\usepackage{amssymb}
\usepackage{float}
\begin{document}

\title{Skyrmion Ratchet in Funnel Geometries}
\author{J. C. Bellizotti Souza$^1$, N. P. Vizarim$^2$, C. J. O. Reichhardt$^3$, C. Reichhardt$^3$ and P. A. Venegas$^1$}
\affiliation{$^1$ Departamento de F\'isica, Faculdade de Ci\^encias, Unesp-Universidade Estadual Paulista, CP 473, 17033-360 Bauru, SP, Brazil\\
$^2$ POSMAT - Programa de P\'os-Gradua\c{c}\~ao em Ci\^encia e Tecnologia de Materiais, Faculdade de Ci\^encias, Universidade Estadual Paulista - UNESP, Bauru, SP, CP 473, 17033-360, Brazil\\
$^3$ Theoretical Division and Center for Nonlinear Studies, Los Alamos National Laboratory, Los Alamos, New Mexico 87545, USA}

\date{\today}

\begin{abstract}
Using a particle-based model, we simulate the behavior of a skyrmion under the influence of asymmetric funnel geometries and ac driving at zero temperature. We specifically investigate possibilities for controlling the skyrmion motion by harnessing a ratchet effect. Our results show that as the amplitude of a unidirectional ac drive is increased, the skyrmion can be set into motion along either the easy or hard direction of the funnel depending on the ac driving direction. When the ac drive is parallel to the funnel axis, the skyrmion flows in the easy direction and its average velocity is quantized. In contrast, when the ac drive is perpendicular to the funnel axis, a Magnus-induced ratchet effect occurs, and the skyrmion moves along the hard direction with a constant average velocity. For biharmonic ac driving of equal amplitude along both the parallel and perpendicular directions, we observe a reentrant pinning phase where the skyrmion ratchet vanishes. For asymmetric biharmonic ac drives, the skyrmion exhibits a combination of effects and can move in either the easy or hard direction depending on the configuration of the ac drives.  These results indicate that it is possible to achieve controlled skyrmion motion using funnel geometries, and we discuss ways in which this could be harnessed to perform data transfer operations. 
\end{abstract}

\maketitle

\vskip 2pc

\section{Introduction}

The flow of particles interacting with asymmetric landscapes can be controlled
by means of a ratchet effect, which is a net dc transport that
emerges from a combination of broken spatial symmetry
and an oscillating drive. The ratchet effect can 
also be understood in terms of a diode effect,
where the asymmetry produces different depinning forces in different
directions, 
yielding a preferential or ``easy'' direction of motion
\cite{reimann_brownian_2002}. This effect has been
investigated in several systems such as protein motors
\cite{vale_protein_1990, hwang_structural_2019, allen_two-way_2016}, molecular motors
\cite{ait-haddou_brownian_2003,astumian_fluctuation_1994,wang_ratchets_2002}, colloids
\cite{libal_dynamics_2006,lee_observation_2005}, 
type II superconducting vortices
\cite{lee_reducing_1999,wordenweber_guidance_2004,lu_reversible_2007,gillijns_origin_2007,reichhardt_reversible_2015,lin_rectification_2011,reichhardt_commensurability_2010,de_souza_silva_controlled_2006,villegas_experimental_2005,olson_reichhardt_rectification_2005,wambaugh_superconducting_1999,yu_vortex_2010,Yu07,Plourde09,Palau12,Shklovskij11,Dobrovolskiy20,Lyu21}
, electrons \cite{song_electron_2002}, active matter
\cite{reichhardt_active_2013,reichhardt_ratchet_2017,mcdermott_collective_2016,leonardo_bacterial_2010} 
and recently in skyrmions \cite{reichhardt_magnus-induced_2015,chen_skyrmion_2019,chen_ultrafast_2020,ma_reversible_2017,vizarim_skyrmion_2020,vizarim_guided_2021,gobel_skyrmion_2021}.
Reversals of motion from the easy to the hard direction may occur as a 
function of the applied magnetic field or other variables \cite{lu_reversible_2007,de_souza_silva_controlled_2006,villegas_experimental_2005,olson_reichhardt_rectification_2005,Shklovskij14,Dinis07,PerezdeLara10,PerezdeLara11} due to collective interactions 
between particles. One of the earliest proposals of a
two-dimensional (2D) periodic asymmetric potential that can induce a
ratchet effect was the asymmetric
funnel geometry \cite{wambaugh_superconducting_1999,Reichhardt10a}. 
Wambaugh {\it et al.} used a combination of periodic
asymmetric funnels and 
ac driving to produce a net dc motion of type II superconducting vortices.
Advanced sample fabrication techniques made it possible to
produce this type of vortex ratchet experimentally
\cite{yu_vortex_2010,VlaskoVlasov13,Karapetrov12}.
Most of the work on ratchets constructed from asymmetric potentials involves
overdamped particles such as colloids or vortices.
Recently, however, there has been great
interest in the dynamics of skyrmions and proposals to control their motion,
and a ratchet effect could be an ideal method for achieving such control.

Magnetic skyrmions are spin textures pointing in all directions
and wrapping a sphere in order to form a topologically stable 
particle-like object \cite{nagaosa_topological_2013}
which
can be set into motion by the application of a spin polarized current
\cite{jonietz_spin_2010,schulz_emergent_2012,yu_skyrmion_2012,iwasaki_current-induced_2013,lin_driven_2013}.
In the presence of such
external driving, skyrmions
can exhibit a depinning threshold similar to
that found for superconducting vortices, and
it is possible to perform
transport measurements and construct
skyrmion velocity-force curves
\cite{schulz_emergent_2012,iwasaki_current-induced_2013,lin_driven_2013,liang_current-driven_2015,woo_observation_2016}. 
Skyrmions show huge promise for applications in 
spintronics such as race-track memory devices due to
their small size and their ability to be displaced by
low currents 
\cite{fert_skyrmions_2013,fert_magnetic_2017,sampaio_nucleation_2013}.
In order to realize devices of this type,
new methods must be developed which permit the skyrmion motion 
to be controlled precisely.

A key difference between skyrmions and overdamped particles is that the
skyrmion dynamics is dominated by the Magnus term
\cite{nagaosa_topological_2013,schulz_emergent_2012,iwasaki_current-induced_2013,fert_magnetic_2017,muhlbauer_skyrmion_2009},
which is negligible in other systems.
The
Magnus term produces a skyrmion velocity contribution
that is perpendicular to the applied external drive.
It has been proposed that the Magnus force is responsible for the
low value of the skyrmion depinning threshold current
\cite{iwasaki_current-induced_2013,lin_particle_2013}. 
In the absence of pinning, the skyrmion moves
at an angle with respect to the driving direction known as the
intrinsic skyrmion Hall angle, $\theta_{sk}^{\rm int}$ \cite{nagaosa_topological_2013,fert_magnetic_2017,jiang_direct_2017}. 
This angle depends on the ratio between the Magnus term, $\alpha_{m}$,
and the damping term, $\alpha_{d}$. 
Depending on the system parameters,
this ratio
can be small or very high
\cite{nagaosa_topological_2013,jiang_direct_2017,litzius_skyrmion_2017,woo_current-driven_2018,juge_current-driven_2019,zeissler_diameter-independent_2020}.

The earliest proposal for a skyrmion ratchet
\cite{reichhardt_magnus-induced_2015}
involved
the same type of quasi-one-dimensional (1D) asymmetric substrate
that was known to produce ratcheting of superconducting vortices.
For an ac drive applied parallel to the substrate asymmetry
direction, the skyrmion exhibits a quantized net 
displacement,
while a Magnus-induced transverse ratchet
effect appears when the ac drive is applied 
perpendicular to the substrate asymmetry direction.
Further work demonstrated the creation of
a vector ratchet with a direction of motion that can be rotated
controllably with respect to
the orientation of the
substrate asymmetry
\cite{ma_reversible_2017}. Recently, a
skyrmion ratchet effect produced by a combination of spatial asymmetry and
ac driving
was used to guide
the skyrmion motion in racetrack storage devices 
\cite{gobel_skyrmion_2021}.
Other skyrmion ratchet effects have been proposed based on
currents with a loop structure \cite{yamaguchi_control_2020} or
biharmonic in-plane magnetic fields \cite{chen_skyrmion_2019}.

Using numerical simulations, we investigate the skyrmion behavior in an
asymmetric funnel geometry under applied ac
driving at zero temperature.
We find
that when an ac drive is applied parallel to the funnel axis, 
a skyrmion ratchet effect occurs in which the skyrmions
move with quantized velocities along the easy direction.
On the other hand, 
when the ac drive is applied perpendicular to the funnel axis,
there is a Magnus-induced transverse skyrmion 
ratchet effect in which the skyrmion moves along the hard direction
at a constant average 
velocity.
The ability to control the direction of motion of the
skyrmion ratchet by varying the ac driving direction can be useful for 
spintronics applications such as data transfer.
To explore this possibility, we simulate a situation in which
the skyrmion moves along the easy direction for a period of time,
and then switches to motion along the hard direction when the ac driving is
rotated by 90$^\circ$.

\section{Simulation}

\begin{figure}[h]
    \centering
    \includegraphics[width=\columnwidth]{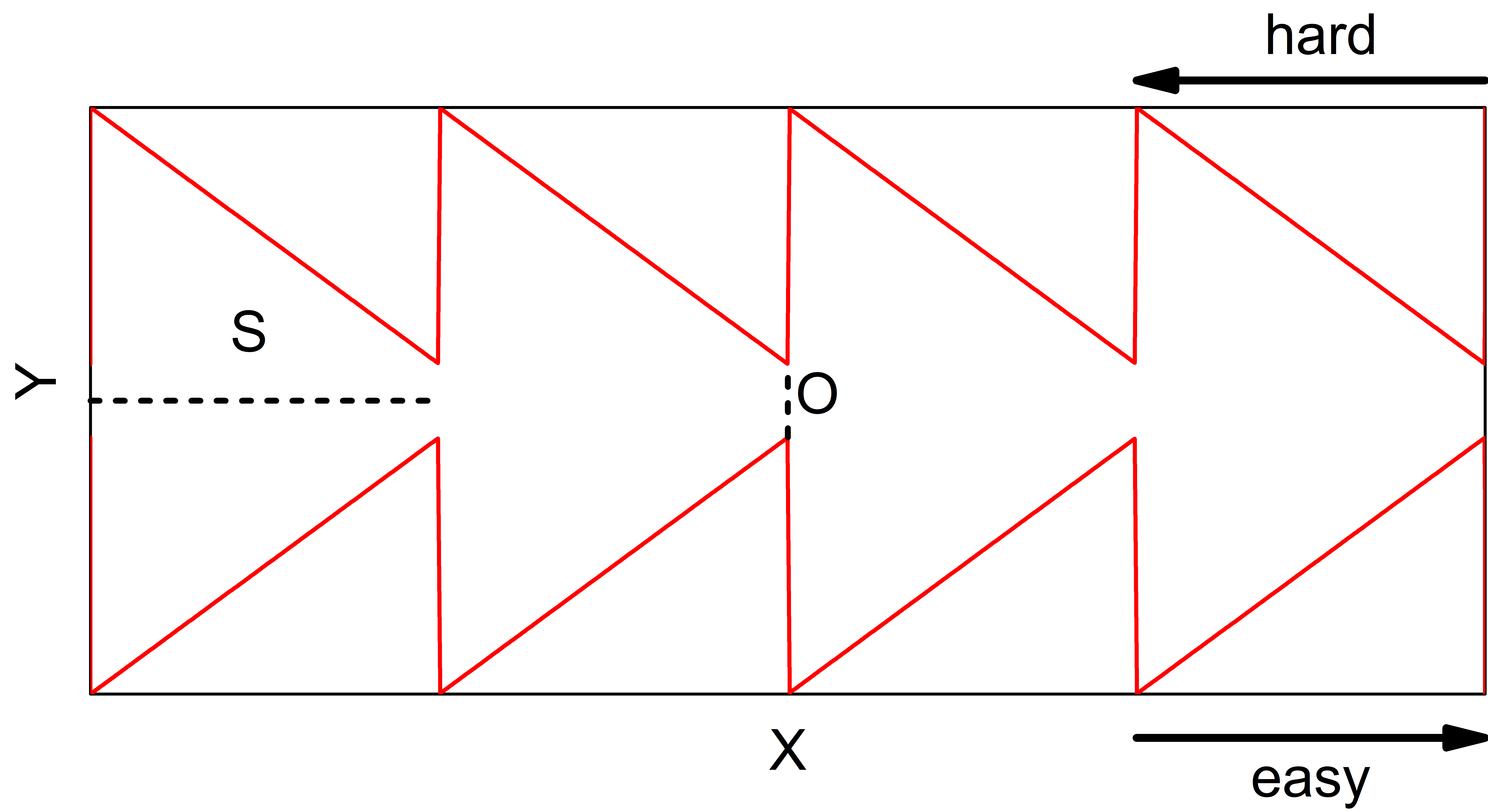}
    \caption{Illustration of the funnel geometry used in this work,
      where $S$ is the length of the funnel and
      $O$ is the width of the funnel opening.
      In this case the number of funnels $N_F=4$, $O=1.0$,
      and the easy and hard directions of motion are labeled.}
    \label{Fig1}
\end{figure}

We simulate a single skyrmion in an $L\times H$ two-dimensional
box with periodic boundary 
conditions in the $x$ direction,
as illustrated in Fig.~\ref{Fig1}.
The skyrmion interacts with an asymmetric funnel 
array aligned with the $x$ direction,
where $N_F$ is the number of funnels, $S$
is the length of each funnel, and $O$ is the width of the
funnel opening.
The asymmetry permits ``easy'' flow along
the positive
$x$ direction and
``hard'' flow along
the negative
$x$ direction.
The skyrmion dynamics is governed by
the following particle based equation of
motion \cite{lin_particle_2013}:
\begin{equation} \label{eq:1}
      \alpha_d\mathbf{v}+\alpha_m\hat{z}\times\mathbf{v}=\mathbf{F}^{W}+\mathbf{F}^{AC}
\end{equation}
where {\bf v} is the skyrmion velocity.
The first term on the left hand side
containing the damping constant $\alpha_d$
represents the damping
originating from the spin precession and dissipation of electrons
in the skyrmion core.
The second term on the left hand side is
the Magnus force, where $\alpha_m$ is the Magnus constant.
On the right hand side, $\mathbf{F}^{W}$ represents the force exerted by
the confining funnel wall.
The funnel outline is formed by
two asymmetric sawtooth functions and the wall potential is given by
a Gaussian form,
$U\left(\mathbf{r}_{iw}\right)=U_0e^{-\left(\frac{r_{iw}}{a_0}\right)^2}$, where 
$\mathbf{r}_{iw}$ is the shortest distance between the skyrmion and the wall,
$U_0$ is the potential 
strength, and $a_0$ is the wall thickness.
The resulting force is $\mathbf{F}^W=-\nabla U_0 = 
-F_0r_{iw}e^{-\left(\frac{r_{iw}}{a_0}\right)^2}\mathbf{\hat{r}}_i$, where $F_0=\frac{2U_0}{{a_0}^2}$. The
term $\mathbf{F}^{AC}$ represents the applied ac drive,
$\mathbf{F}^{AC}=A\sin\left(2\pi\omega t\right)\hat{\mathbf{x}}+B\cos\left(2\pi\omega 
t\right)\hat{\mathbf{y}}$,
where $\omega$ is the frequency of the ac drive and
$A$ and $B$ are the ac drive amplitudes in the $x$ and $y$
directions, respectively.
For each value of $A$ or $B$ we perform a simulation spanning
$5\times 10^6$ time steps,
with a step size of $10^{-3}$,
in order to ensure the system has reached
a steady state.
We measure the 
average velocities $\left\langle V_x\right\rangle=2\pi \left\langle \mathbf{v} \cdot \widehat{\rm 
{\bf{x}}}\right\rangle / \omega S$ and $\left\langle V_y\right\rangle = 2\pi \left\langle \mathbf{v} \cdot
\widehat{\rm {\bf{y}}}\right\rangle / \omega S$,
where $S$ is the length of a single funnel.
The value of $A$ or $B$ is incremented by $\delta A(B)=0.005$.
In this work we fix $U_0=1.0$, 
$\omega=2\times 10^{-5}$, the sample length $L=20.0$,
the sample height $H=8.0$, and $a_0=0.02$.
We normalize all distances by
the screening length $\xi$ and select
the damping and Magnus constants
such that ${\alpha_m}^2+{\alpha_d}^2=1$.

\section{AC drive in the $\mathbf{x}$ direction}

\begin{figure}[h]
   \centering
   \includegraphics[width=0.9\columnwidth]{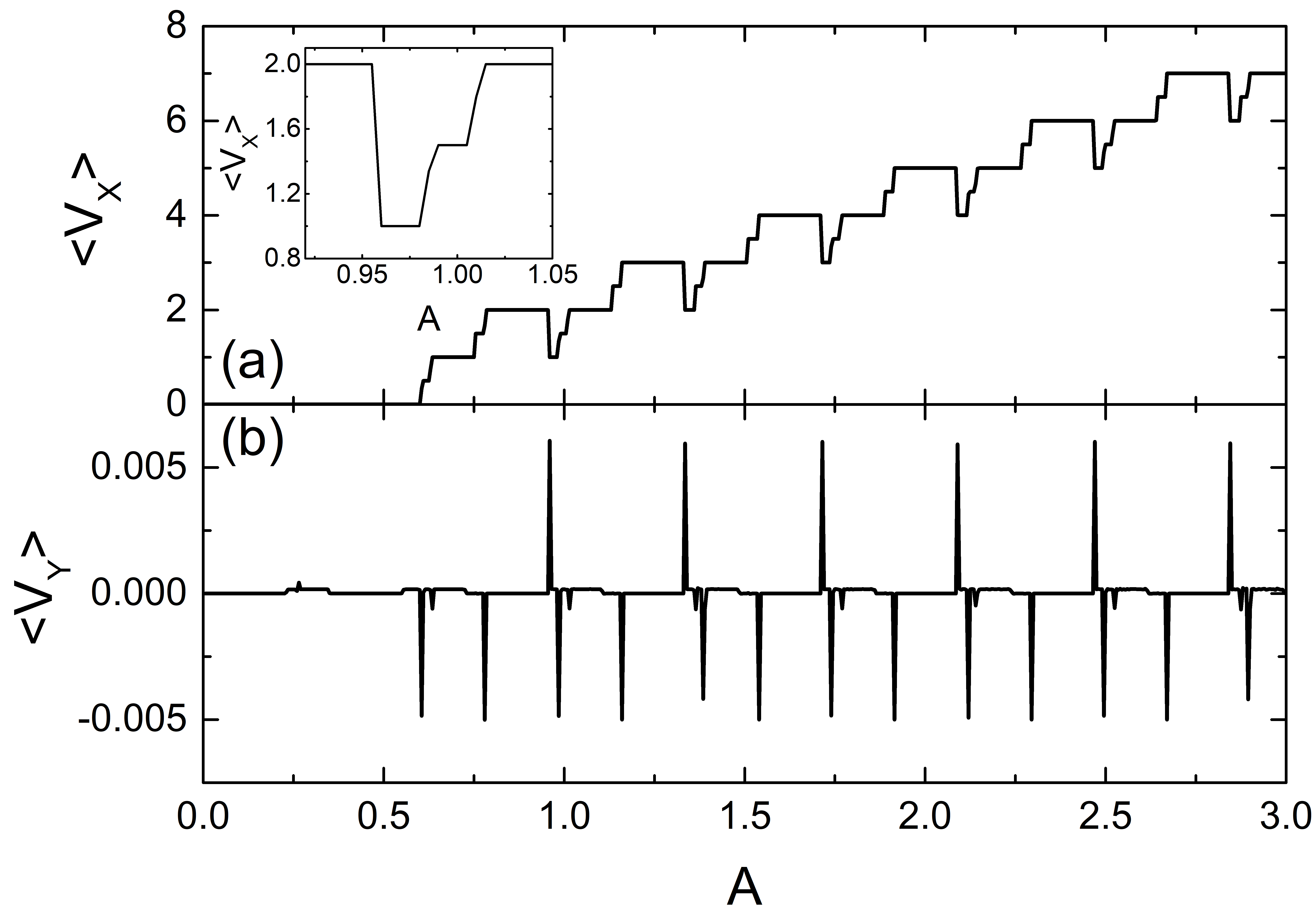}
   \caption{(a) $\langle V_x\rangle$ and (b) $\langle V_y\rangle$
     versus ac drive amplitude $A$ for a
     sample with $\alpha_m/\alpha_d = 0.5$, $N_F=4$, and $O=1.0$.
     Here the ac driving is only along the $x$ direction.
     The inset of (a) shows a blowup of panel (a)
     over the range $0.92<A<1.05$.
 }
   \label{Fig2}
\end{figure}
    
We first consider an ac drive applied parallel
to the funnel array axis, where
$A\neq 0$ and $B=0$.
In Fig.~\ref{Fig2} we plot $\langle V_x\rangle$ and $\langle V_y\rangle$ as a 
function of the ac drive amplitude $A$ for
a system with $\alpha_m/\alpha_d = 0.5$.
The average $x$ velocity $\langle V_x\rangle$,
shown in
Fig.~\ref{Fig2}(a),
has a step-like behavior superimposed on a monotonic increase.
Here the combination of the ac
drive and the broken spatial symmetry of the funnel array
generates a ratchet 
effect in which there is a net dc motion of the skyrmion
in the $+x$ direction. Although $\langle V_y\rangle$
in Fig.~\ref{Fig2}(b)
is very close to zero, there is a spike of transverse
motion at the edge of every step in $\langle V_x\rangle$.

\begin{figure}[h]
  \centering
  \includegraphics[width=1.0\columnwidth]{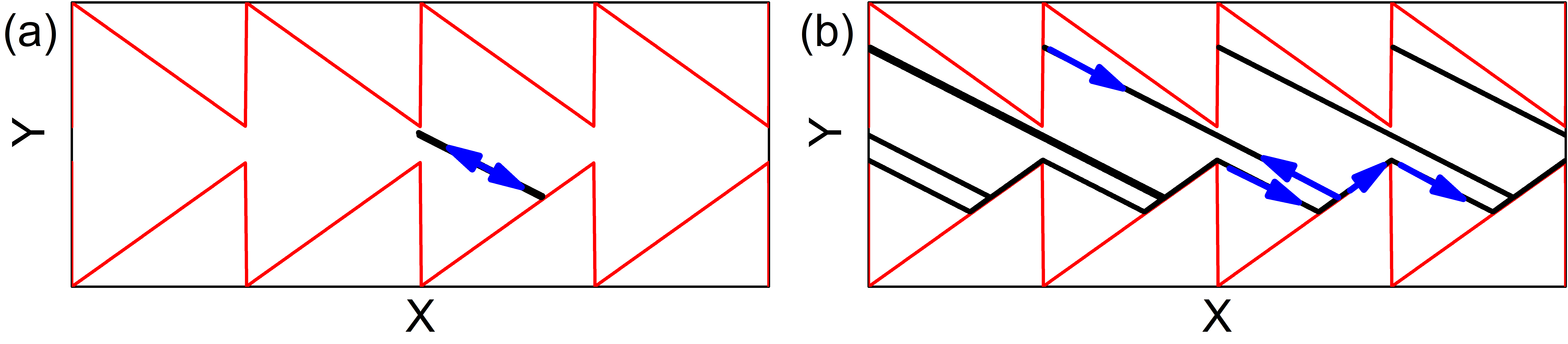}
  \includegraphics[width=1.0\columnwidth]{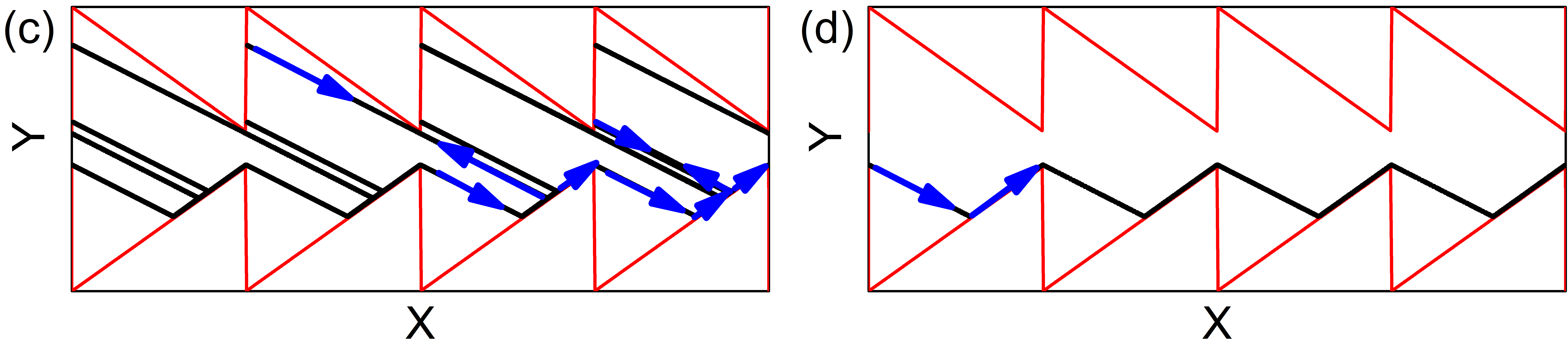}
\caption{Funnel wall (red lines) and the skyrmion trajectory (black lines)
for a single skyrmion interacting with a funnel array
with $N_F=4$, $O=1.0$, $S=5$,
$\alpha_m/\alpha_d = 0.5$, and an ac drive applied
along the $x$ axis with $B=0$.
Blue arrows indicate the direction of skyrmion motion.
(a)
At $A=0.25$, the skyrmion forms a closed orbit
inside a single funnel.
(b)
At $A=0.97$, there is a translating orbit with a net dc
motion along the $+x$ direction. On average the skyrmion translates by a distance of
 $1.0S$ per ac drive cycle.
(c)
At $A=1.76$, the skyrmion translates
an average of $3.5S$
per cycle.
(d) At $A=2.4$ the skyrmion
translates an average
of $6S$ per cycle.
}
    \label{Fig3}
\end{figure}    

In Fig.~\ref{Fig3} we illustrate some representative
skyrmion trajectories for the system in Fig.~\ref{Fig2}.
At $A=0.25$, Fig.~\ref{Fig3}(a) shows that the ac drive 
amplitude is not large enough to generate
a ratchet motion and the skyrmion forms a localized orbit
inside one of the funnels.
For $A=0.97$ in Fig.~\ref{Fig3}(b),
we find $+x$ direction skyrmion ratchet motion
in which the skyrmion
translates by an average distance of $1.0S$
during each ac drive
cycle.
The trajectory trace indicates that
the skyrmion moves forward by $2S$
during the $+x$ portion of the
ac drive cycle, and
backward by
$1.0S$
during the $-x$ portion of the drive cycle.
In Fig.~\ref{Fig3}(c) at $A=1.76$, the skyrmion
translates an average of $3.5S$ per cycle,
while at $A=2.4$ in Fig.~\ref{Fig3}(d), the skyrmion flows
faster at a rate of $6S$ per cycle,
always 
in the $+x$ direction.
    
\begin{figure}[h]
    \centering
    \includegraphics[width=1.0\columnwidth]{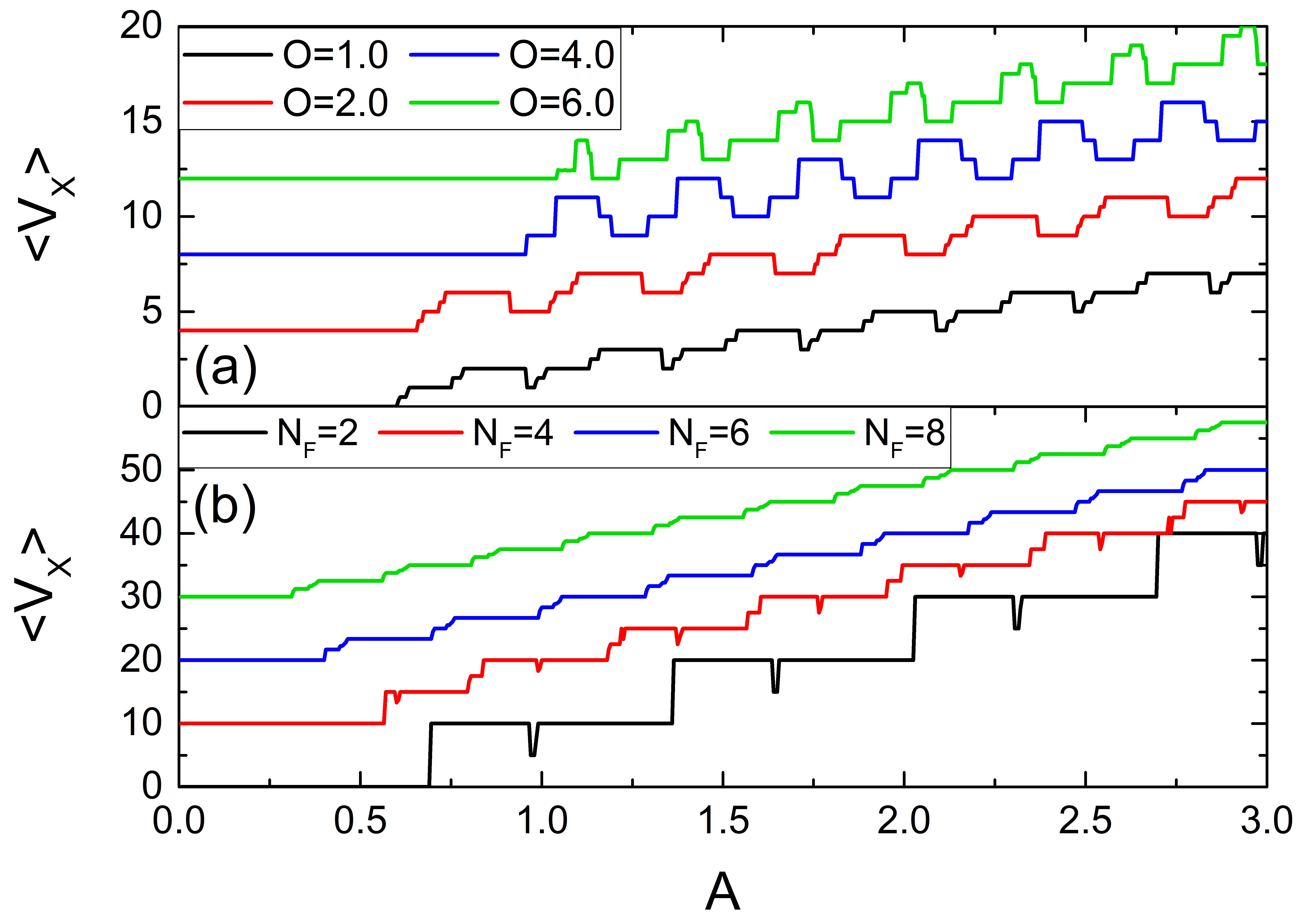}
    \caption{
Plots of $\langle V_x\rangle$ vs $A$
in samples with $\alpha_m/\alpha_d=0.5$.
(a) Varied funnel opening widths $O$ for a fixed number of funnels $N_F=4$.
(b) Varied $N_F$ for fixed $O=0.28$.
For clarity, the curves have been offset vertically.
To facilitate comparison, all velocities are normalized using 
$S=5$ even though $S$ varies when $N_F$ is changed.
    }
    \label{Fig4}
\end{figure}

\begin{figure}[h]
    \centering
    \includegraphics[width=1.0\columnwidth]{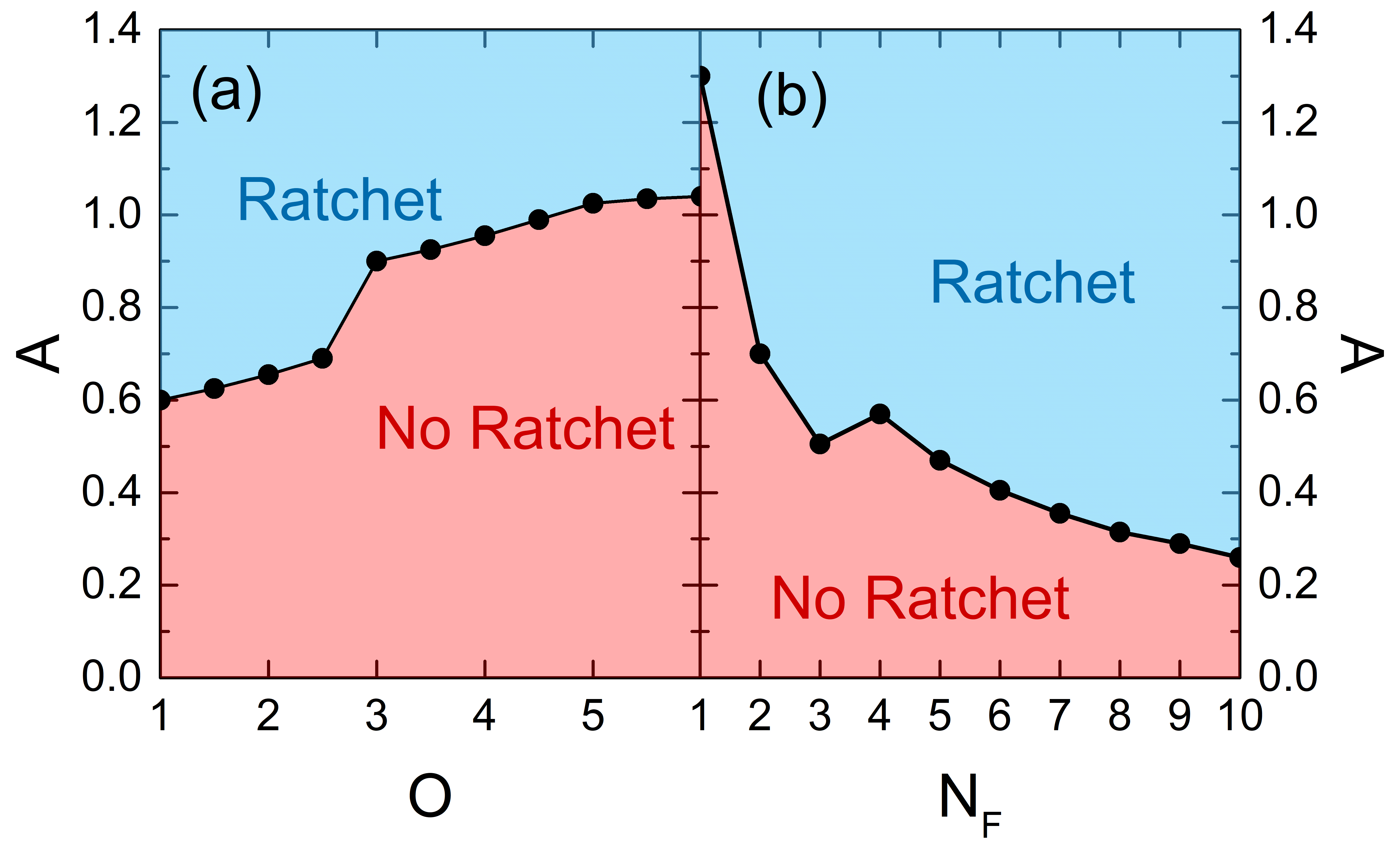}
    \caption{
Dynamic phase diagram as a function of      
ac drive amplitude $A$ vs (a) $O$ for $N_F=4$ and (b) $N_F$ for $O=0.28$
in a system with $\alpha_m/\alpha_d=0.5$.
Red regions exhibit no ratchet effect, while blue regions indicate the appearance
of ratcheting in the $+x$ direction.
    }
    \label{Fig5}
\end{figure}

Since the ratchet effect requires asymmetry in the potential,
we expect that
changes to the funnel 
shape or density will alter the skyrmion behavior.
In Fig.~\ref{Fig4}(a) we show the results of changing the width $O$ of the
funnel opening while holding the number of funnels fixed
at $N_F=4$,
while in Fig.~\ref{Fig4}(b) we vary $N_F$ and fix $O=0.28$.
Figure~\ref{Fig4}(a)
shows that when the funnel opening width $O$ is reduced,
the onset of the ratchet effect shifts to lower values of $A$.
This is due to a reduction in the probability that the skyrmion will
be able to pass backwards through the funnel during the $-x$ portion of
the drive cycle as the funnel opening becomes narrower.
We also find that 
the shape of the $\langle V_x\rangle$ signal is modified as $O$ is varied.
In Fig.~\ref{Fig4}(b), where $O$ is fixed to $O=0.28$,
the ratcheting effect is enhanced as the number of funnels
$N_F$ increases.
Here, the length $S$ of each funnel diminishes as $N_F$ becomes larger since
the width of the sample is held constant. As a result, the distance the
skyrmion must travel to move from one funnel plaquette to the next
decreases with increasing $N_F$, and the amplitude $A$ at which the ratchet
effect first appears shifts to lower values.
To summarize
the influence of the funnel shape on
the skyrmion behavior,
we plot dynamic phase diagrams as a function of ac drive amplitude $A$ versus
$O$ in Fig.~\ref{Fig5}(a) and versus $N_F$ in Fig.~\ref{Fig5}(b).
Figure~\ref{Fig5}(a)
shows that the onset of ratchet motion drops to smaller $A$ as the
funnel opening width $O$ is reduced, while
in Fig.~\ref{Fig5}(b),
the threshold amplitude for ratchet motion generally
diminishes as
the number of funnels
increases.
    
\begin{figure}[h]
    \centering
    \includegraphics[width=1.0\columnwidth]{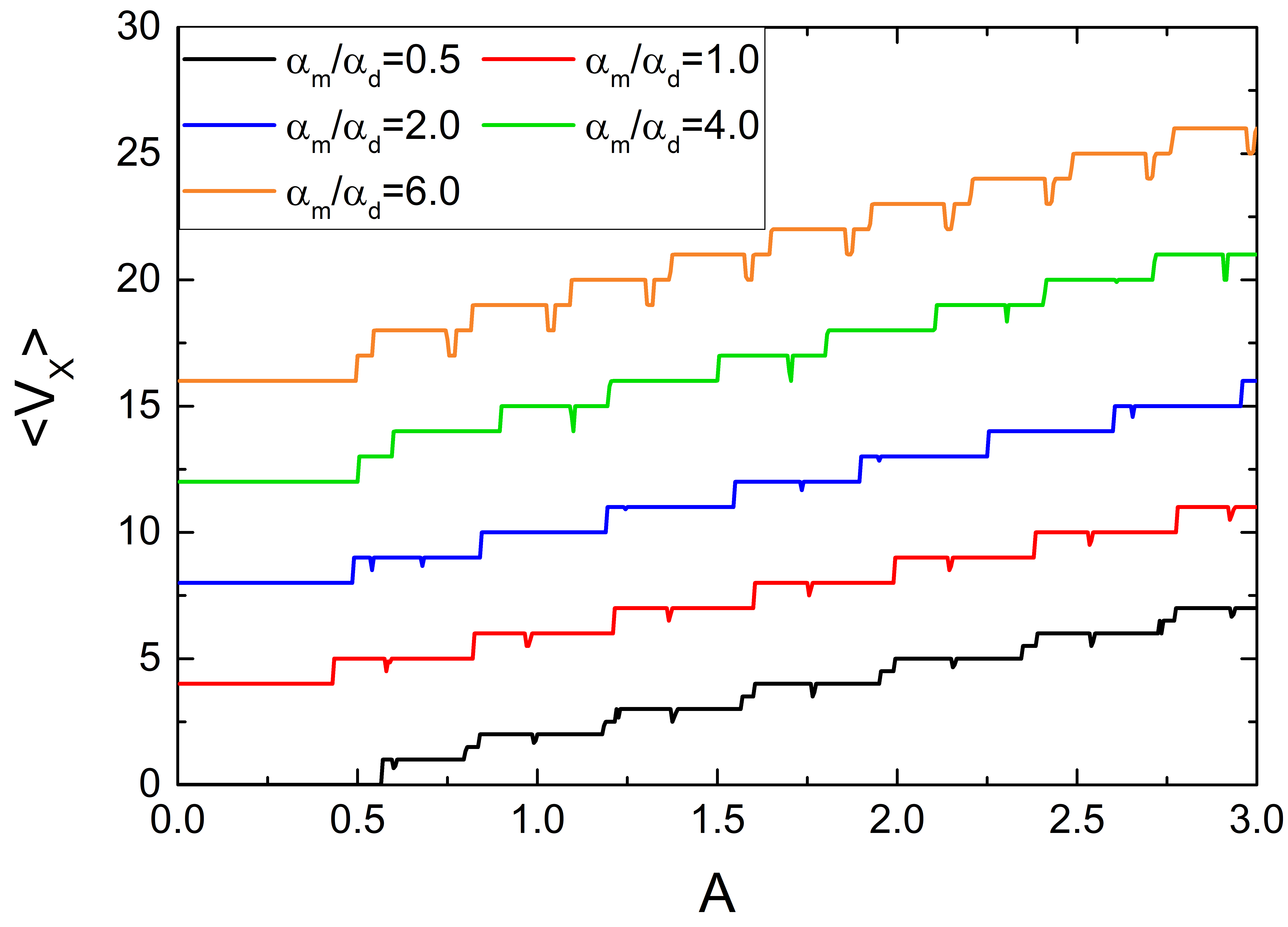}
    \caption{
$\langle V_x\rangle$ vs $A$ for different values of $\alpha_m/\alpha_d$
in samples with $O=0.28$ and $N_F=4$. The curves have been offset vertically
for clarity.     
    }
    \label{Fig6}
\end{figure}

To explore the effect of the Magnus term intensity
on the dynamics,
in Fig.~\ref{Fig6} we plot $\langle V_x\rangle$ versus $A$
at different values of $\alpha_m/\alpha_d$ ranging from $\alpha_m/\alpha_d=0.5$
to $\alpha_m/\alpha_d=6.0$ for a system with $O=0.28$ and $N_F=4$.
As the Magnus term becomes
stronger, the skyrmion velocity steps become narrower, while there is
little change in the depinning threshold.
In previous work on skyrmion ratchets in quasi-1D asymmetric substrates 
\cite{reichhardt_magnus-induced_2015},
when the ac drive was applied parallel to the asymmetric 
potential, increasing the Magnus term
decreased the range of ac driving forces over which the
ratchet motion appeared.
For the funnel geometry considered here,
we find only a weak impact of the Magnus term on the magnitude of the
ratcheting for ac driving applied parallel to the funnel asymmetry direction.

\section{AC drive in the $\mathbf{y}$ direction}

\begin{figure}[h]
    \centering
    \includegraphics[width=0.88\columnwidth]{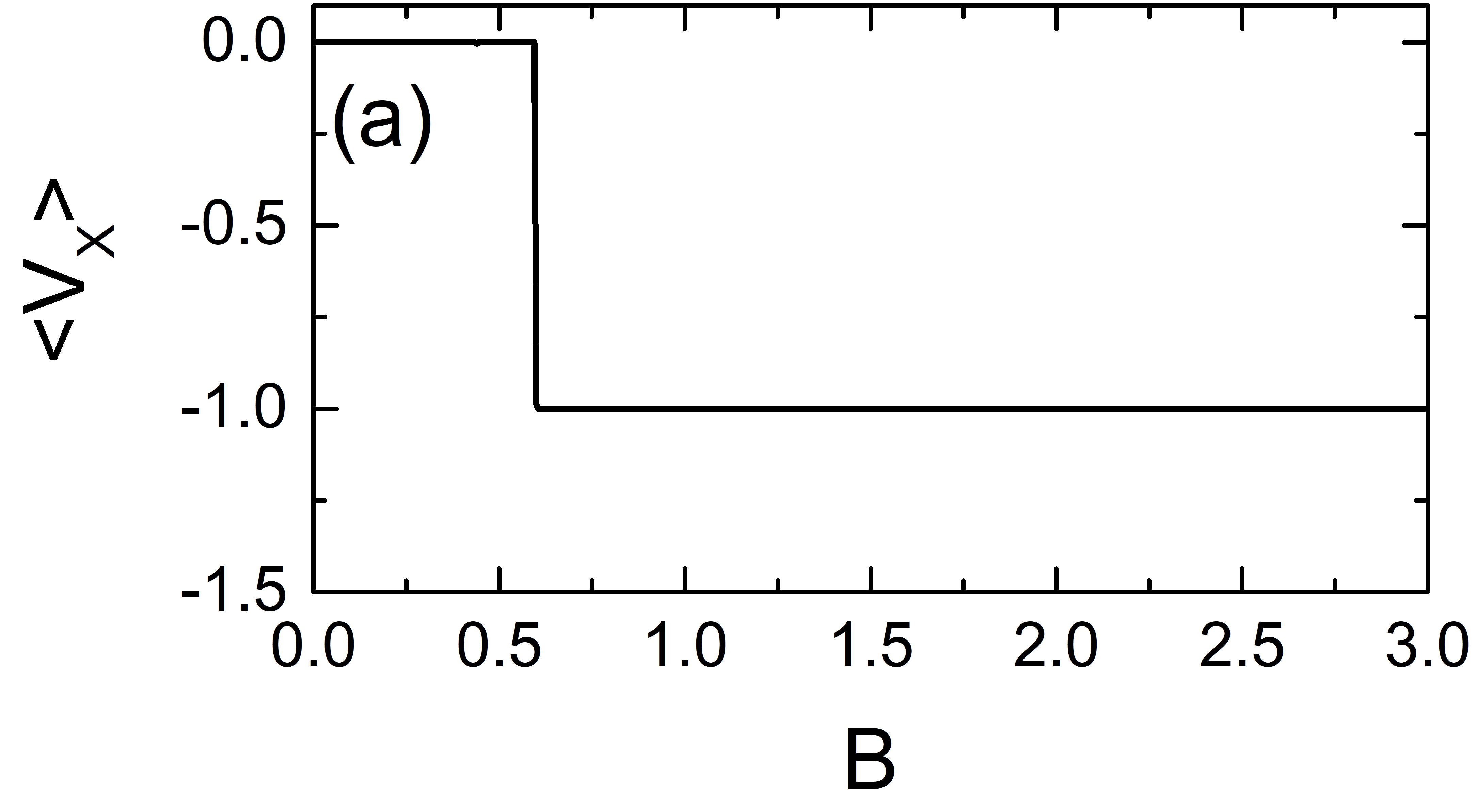}
    \includegraphics[width=0.8\columnwidth]{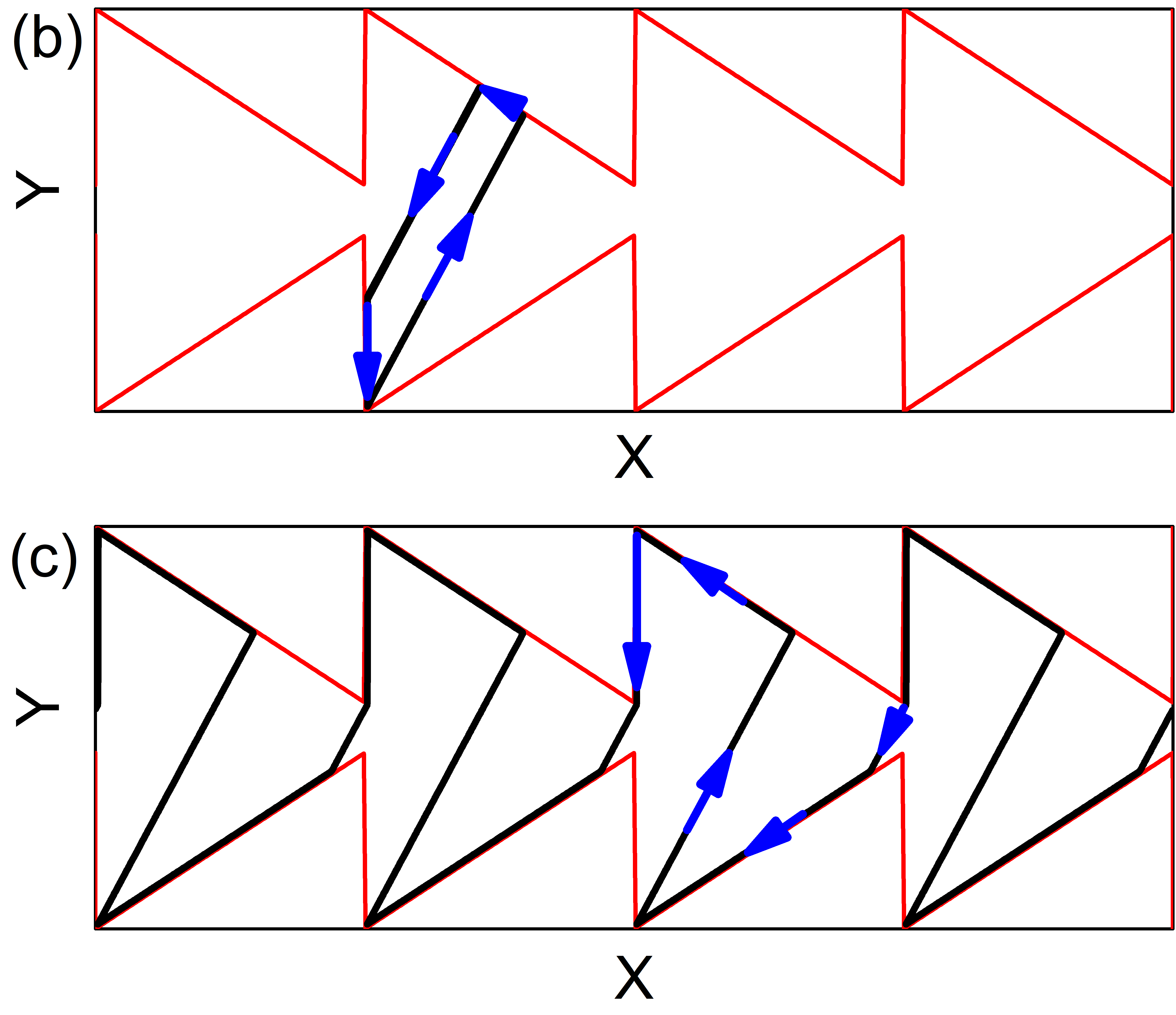}
   \caption{
A system with $N_F=4$, $O=1.0$, $S=5$, $\alpha_m/\alpha_d=0.5$, and an ac
drive applied along the $y$ axis with $A=0$.
(a) $\langle V_x\rangle$ vs ac drive amplitude $B$.
For $B>0.595$, ratcheting motion in the $-x$ direction appears.
(b,c) Funnel wall (red lines) and the skyrmion trajectory (black lines) for
the same system. Blue arrows indicate the direction of skyrmion motion.
(b) At $B=0.5$, the skyrmion forms a stationary orbit localized inside a
single funnel.
(c) Ratcheting motion in the $-x$ direction at $B=1.0$, where
the average velocity is $-1.0 S$ per cycle.
    }
    \label{Fig7}
\end{figure}

We next
consider an ac drive applied perpendicular to the funnel 
axis, with finite $B$ and $A=0$.
In Fig.~\ref{Fig7} we show results from a system with $N_F=4$, $O=1.0$, and 
$\alpha_m/\alpha_d=0.5$.
For 
$B>0.595$, a Magnus-induced transverse ratchet occurs and the skyrmion flows
in the $-x$ direction, as indicated in Fig.~\ref{Fig7}(a).
Note that the motion maintains a steady average velocity of
$\langle V_x\rangle=-1$
over a wide range of applied ac drive amplitudes.
The stability in this motion is of interest
for technological applications since the
translation speed remains the same for various 
values of ac amplitudes.
In Fig.~\ref{Fig7}(b) we illustrate the
localized skyrmion orbit at $B=0.5$, where the 
skyrmion remains trapped inside a single funnel.
In Fig.~\ref{Fig7}(c) we show a skyrmion
ratchet trajectory, where the funnel walls guide the skyrmion motion in
the $-x$ direction.

\begin{figure}[h]
    \centering
    \includegraphics[width=1.0\columnwidth]{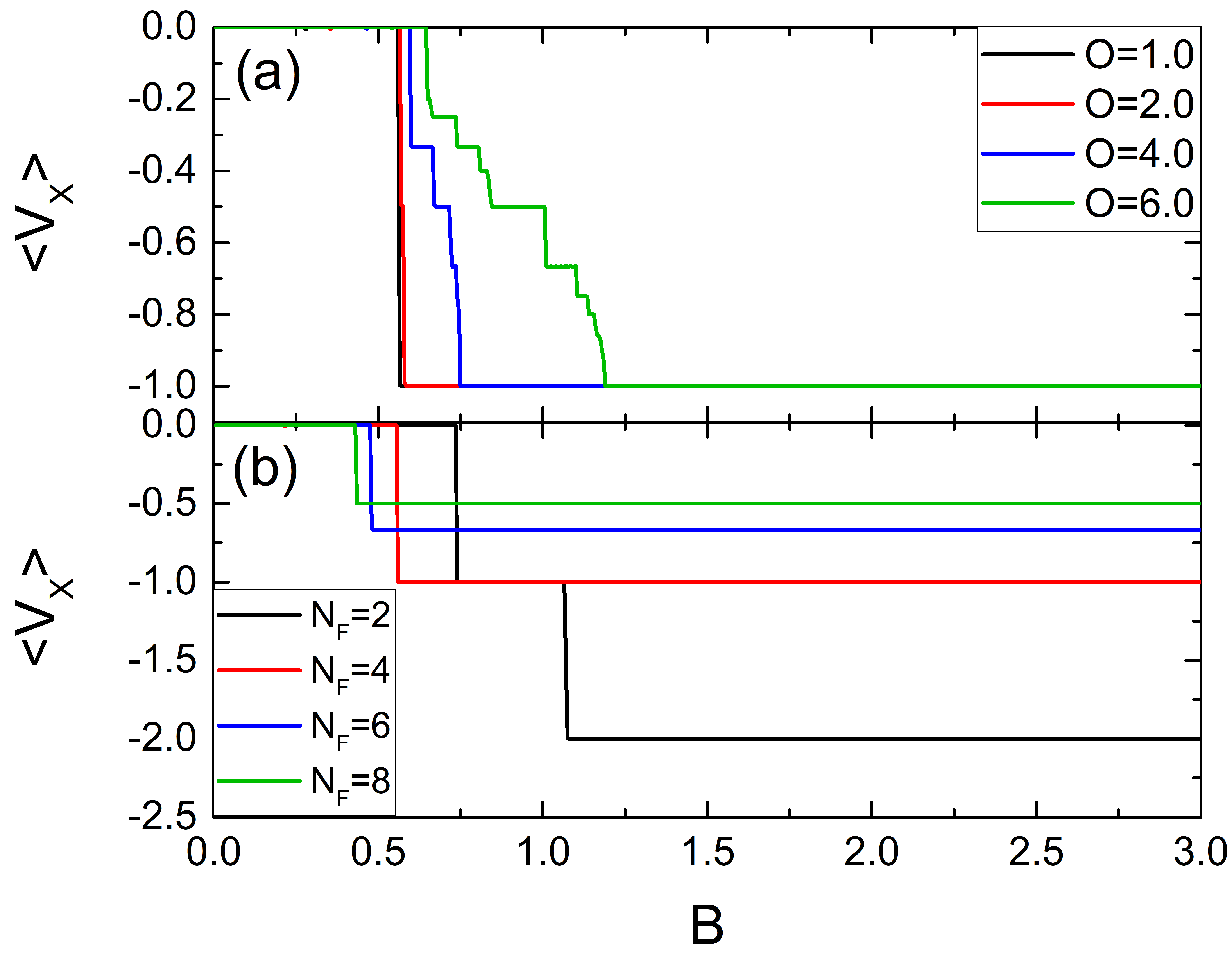}
    \caption{
$\langle V_x\rangle$ vs $B$ for samples with $\alpha_m/\alpha_d=0.5$ and
      $y$ direction ac driving with $A=0$.
      (a) Varied funnel opening widths $O$ and fixed
$N_F=4$. (b) Varied $N_F$ and fixed $O=0.28$. To facilitate
comparison, all velocities are normalized using $S=5$ even though $S$
varies when $N_F$ is changed.
    }
    \label{Fig8}
\end{figure}

\begin{figure}[h]
   \centering
   \includegraphics[width=1.0\columnwidth]{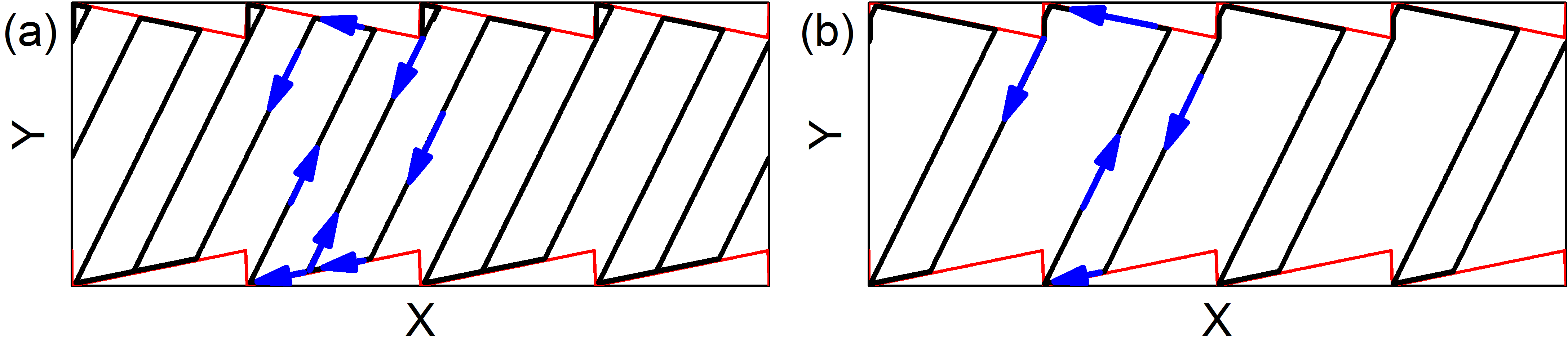}
   \includegraphics[width=1.0\columnwidth]{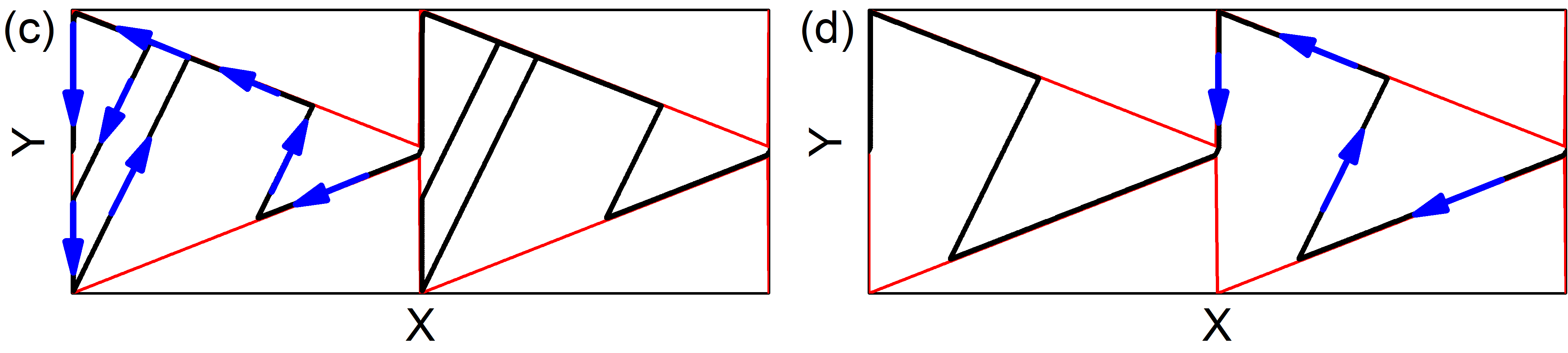}
\caption{Funnel wall (red lines) and the skyrmion trajectory (black lines)
in a system with  
$\alpha_m/\alpha_d = 0.5$ and an ac drive applied along the $y$ axis
with $A=0$,
where a net dc motion along $-x$ appears.
In panels (a) and (b), $N_F=4$ and $O=6.0$, while in
panels (c) and (d), $N_F=2$ and $0=0.28$.
(a) A complex intermediate phase at $B=1.0$.
(b) The saturated stable translating orbit at $B=1.5$.
(c) At $B=1.0$ the net skyrmion speed is reduced.
(d) At $B=1.5$, the skyrmion moves more rapidly 
with $\langle V_x\rangle=-2.0$.
}
\label{Fig9}
\end{figure} 

In Fig.~\ref{Fig8} we show the effect of varying $O$ and $N_F$ on the
net ratcheting motion.
For samples in which the number of funnels is fixed at $N_F=4$ and $O$
is varied over the range $O=1$ to $O=6$, the plot of $\langle V_x\rangle$
versus $B$ in Fig.~\ref{Fig8}(a) 
indicates that there is only a small upward shift in the depinning threshold
as the funnel opening becomes wider.
There is a single sharp transition to motion for small $O$, but at larger $O$
multiple dynamical phases appear and produce numerous steps in the
velocity-force curve.
A variety of skyrmion orbits appear when the funnel openings are wide, as
illustrated in Fig.~\ref{Fig9}(a) at $B=1.0$.
For all values of $O$, there is a critical ac drive amplitude above which
the motion saturates into a state with $\langle V_x\rangle=-1.0$,
as shown in Fig.~\ref{Fig9}(b) at $B=1.5$.
When the number of funnels is increased for fixed $O=0.28$, as in
Fig.~\ref{Fig8}(b), the depinning force decreases since the
skyrmion does not need to move as far to escape from one funnel into
an adjacent funnel.
For this value of $O$, there is only a single translating state except
when $N_F=2$. The larger amount of space inside each funnel for the $N_F=2$
system permits the formation of two different translating motions, as
illustrated in Fig.~\ref{Fig9}(c,d) at $B=1.0$ and $B=1.5$, respectively.

\begin{figure}[h]
    \centering
    \includegraphics[width=1.0\columnwidth]{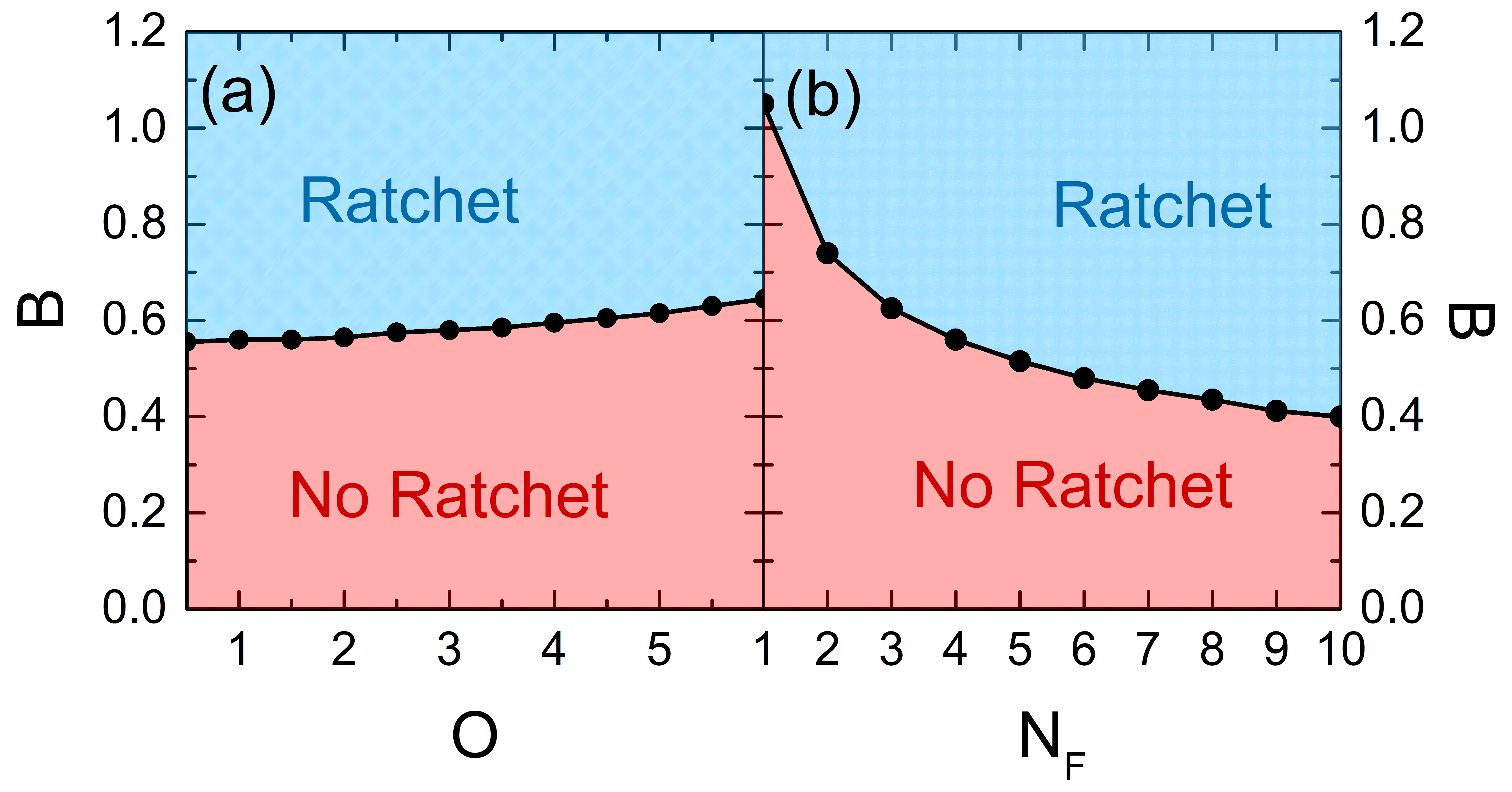}
    \caption{
Dynamic phase diagram as a function of ac drive amplitude $B$ vs      
(a) $O$  for $N_F=4$ and (b) $N_F$ for $O=0.28$ in
a system with $\alpha_m/\alpha_d=0.5$ and $y$ direction ac driving
with $A=0$.
Red regions exhibit no ratchet effect, while blue regions indicate the
appearance of ratcheting in the $-x$ direction.
}
    \label{Fig10}
\end{figure}
    
We summarize
the skyrmion behavior under $y$ direction ac driving in
Fig.~\ref{Fig10} where we plot a dynamic phase diagram as a
function of $B$ versus $O$ for $N_F=4$ in Fig.~\ref{Fig10}(a) and
as a function of $B$ versus $N_F$ for $O=0.28$ in Fig.~\ref{Fig10}(b).
The width of the funnel opening has little effect on the onset of the
ratcheting motion,
as shown in Fig.~\ref{Fig10}(a),
since the skyrmion flow is along the $-x$ direction.
As the number of funnels increases, Fig.~\ref{Fig10}(b) indicates that the
onset of ratcheting motion drops to lower values of $B$.
    
\begin{figure}[h]
    \centering
    \includegraphics[width=1.0\columnwidth]{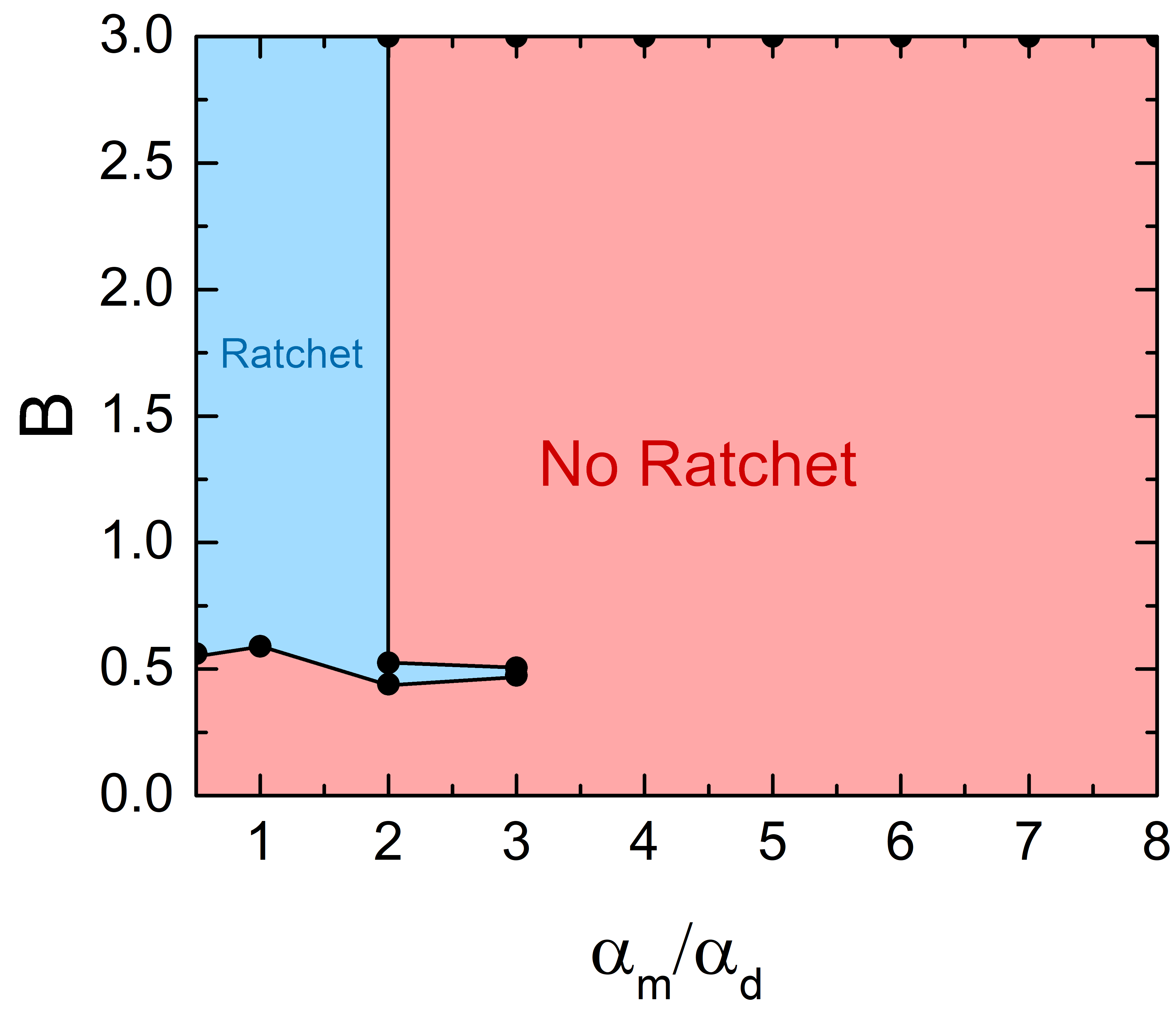}
    \caption{
Dynamic phase diagram as a function of $B$ vs $\alpha_m/\alpha_d$
for samples with $O=0.28$ and $N_F=4$ under $y$ direction ac driving
with $A=0$.
Red regions exhibit no ratchet effect, while blue regions indicate the
appearance of ratcheting in the $-x$ direction.
}
\label{Fig11}
\end{figure}

In Fig. \ref{Fig11}, we plot a dynamic phase diagram as a function
of $B$ versus $\alpha_m/\alpha_d$ for samples with $O=0.28$ and $N_F=4$
to
show how the size of the Magnus term
impacts the transverse ratchet effect.
When $\alpha_m/\alpha_d \geq 3.0$, no ratcheting motion appears.
The reduction of a Magnus-induced transverse 
ratchet as a function of increasing Magnus force
was observed previously in 
systems with quasi-1D substrates \cite{reichhardt_magnus-induced_2015}.
For our funnel geometry,
although Fig.~\ref{Fig6} showed that
$\alpha_m/\alpha_d$ has little impact on
the ratchet effect when the ac drive is applied parallel to the $x$
direction, 
Fig.~\ref{Fig11} indicates that
for ac driving
applied parallel to the $y$ direction,
modifications to the Magnus term can strongly modify the
ratcheting behavior.
For $\alpha_m/\alpha_d=2.0$ and $3.0$,
we observe a reentrant pinning phase with a small window
of ratcheting motion near $B=0.5$.

\section{Biharmonic AC driving}

\begin{figure}[h]
\centering
\includegraphics[width=1.0\columnwidth]{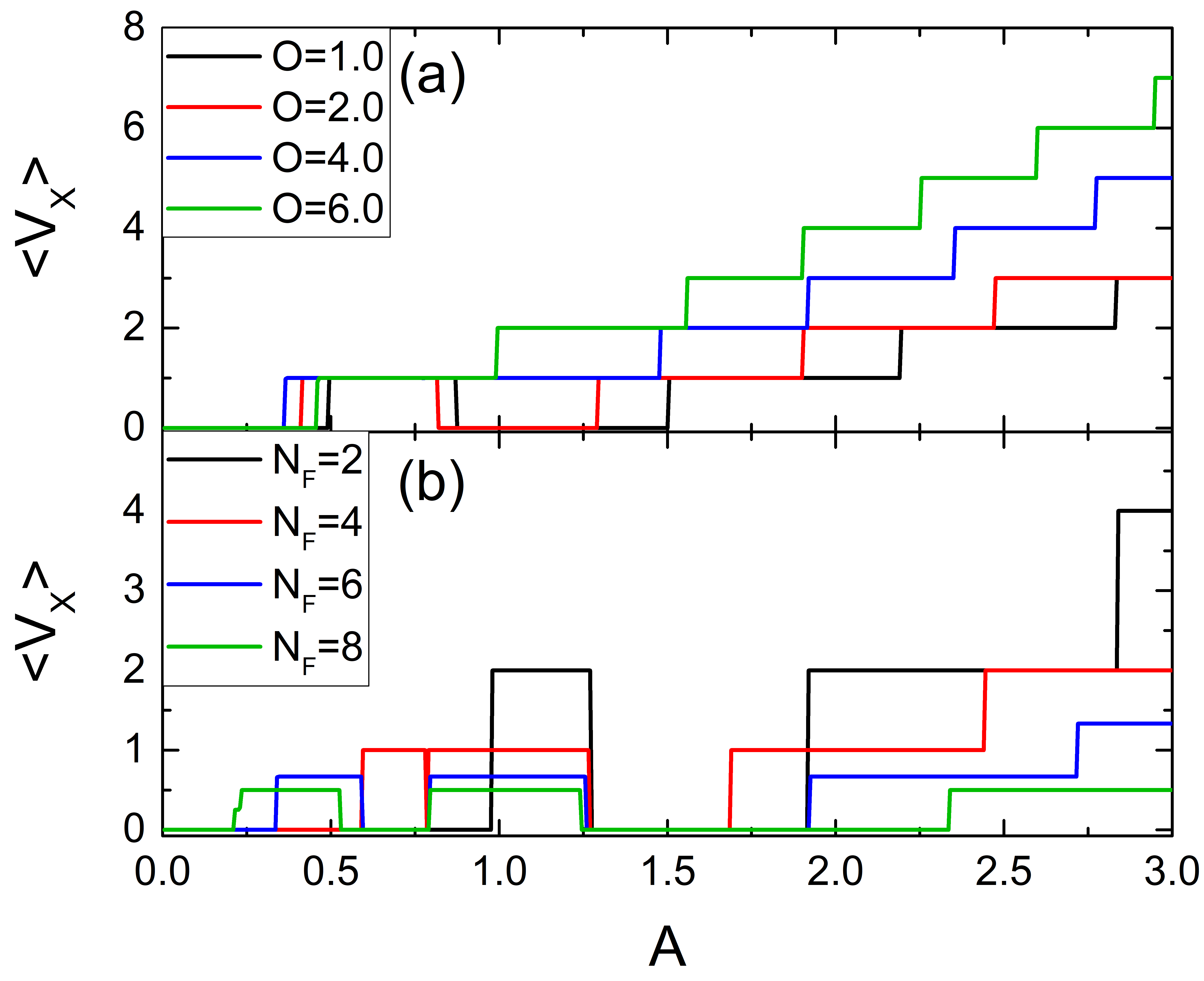}
\caption{
$\langle V_x\rangle$ vs $A$ for samples with biharmonic driving in which
$A=B$ and 
$\alpha_m/\alpha_d=0.5$.
(a) Varied $O$ and fixed $N_F=4$.
(b) Varied $N_F$ and fixed $O=0.28$.
To facilitate comparison, all velocities are normalized using $S=5$ even
though $S$ varies when $N_F$ is changed.
}
\label{Fig12}
\end{figure}

\begin{figure}[h]
    \centering
    \includegraphics[width=1.0\columnwidth]{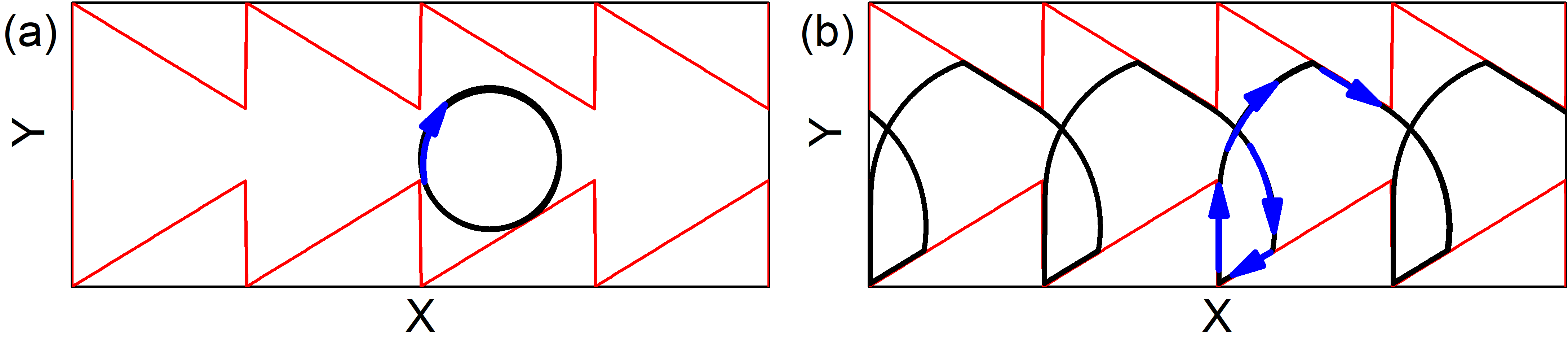}
    \includegraphics[width=1.0\columnwidth]{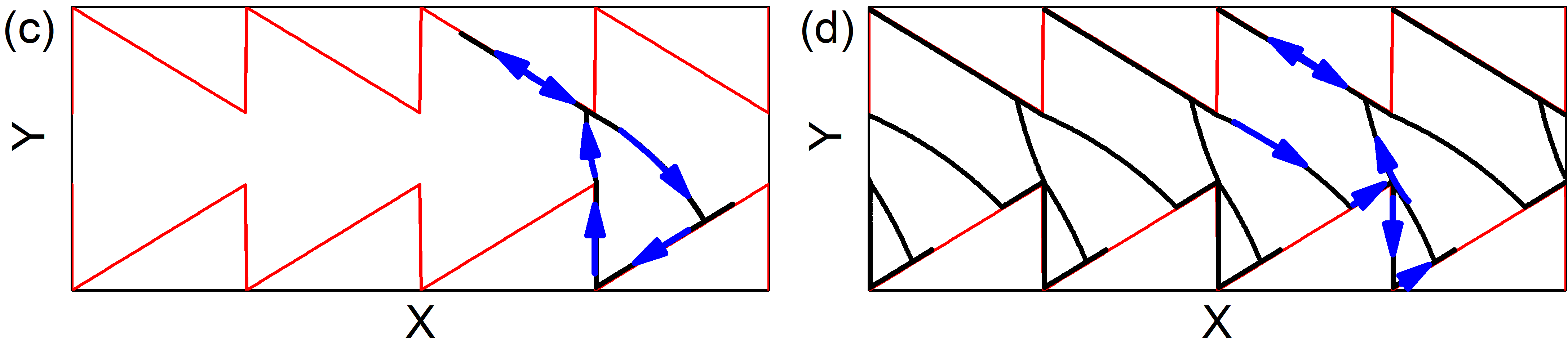}
\caption{Funnel wall (red lines) and the skyrmion trajectory (black lines)
in a system with
$\alpha_m/\alpha_d = 0.5$ and
a biharmonic ac drive applied in both the $x$ and $y$
directions with $B=A$, $N_F=4$ and $O=2.0$. 
(a)
At $A=0.25$, a localized orbit forms and there is no net dc motion.
(b)
At $A=0.5$ the skyrmion translates in the $+x$ direction.
(c)
At $A=1.0$, a complex localized orbit forms between two adjacent funnels
in the reentrant pinning phase.
(d)
At $A=1.5$, above the reentrant pinning window, the skyrmion
has a net motion in the $+x$ direction.
        }
        \label{Fig13}
    \end{figure}

We next consider biharmonic ac driving
where both $A$ and $B$ are finite and the skyrmion
is simultaneously driven in both the $x$ and $y$ directions.
In Fig.~\ref{Fig12} we plot $\langle V_x\rangle$ versus $A$ for
samples with $\alpha_m/\alpha_d=0.5$ and $B=A$.
For a fixed number of funnels $N_F=4$ and varied $O$, as in
Fig.~\ref{Fig12}(a),
a reentrant pinning phase appears
for $O\leq 2.0$ in which the net dc motion disappears.
For higher values of $O$, this reentrant pinning phase vanishes since
multiple trapped orbits can no longer form inside individual funnels.
In Fig.~\ref{Fig12}(b), where we fix $O=0.28$ and vary $N_F$,
we are in a regime where the reentrant pinning phase is always present,
and the range of ac drive amplitudes over which the 
reentrant pinning appears changes with $N_F$.
For large $N_F$, such as
$N_F=6$ and $N_F=8$,
a second reentrant pinning phase emerges.
Here, the shorter funnel lengths at high $N_F$ make it possible for
additional trapped nontranslating skyrmion orbits to form.
In Fig.~\ref{Fig13} we show representative skyrmion trajectories for the system 
from Fig.~\ref{Fig12} with $N_F=4$ and $O=2.0$.
At $A=0.25$ in Fig.~\ref{Fig13}(a), there is no net dc motion
since the drive amplitude is not large enough for a delocalized orbit to form.
In contrast, at $A=0.5$ in Fig.~\ref{Fig13}(b),
we find a translating orbit
with motion in the $+x$ direction.
In Fig.~\ref{Fig13}(c) we illustrate the reentrant pinning phase
at $A=1.0$ where
a localized orbit spanning two funnels appears, while
in Fig.~\ref{Fig13}(d) at $A=1.5$, we find a translating orbit with a net
motion in the $+x$ direction above the reentrant pinning regime.
    
\begin{figure}[h]
    \centering
    \includegraphics[width=1.0\columnwidth]{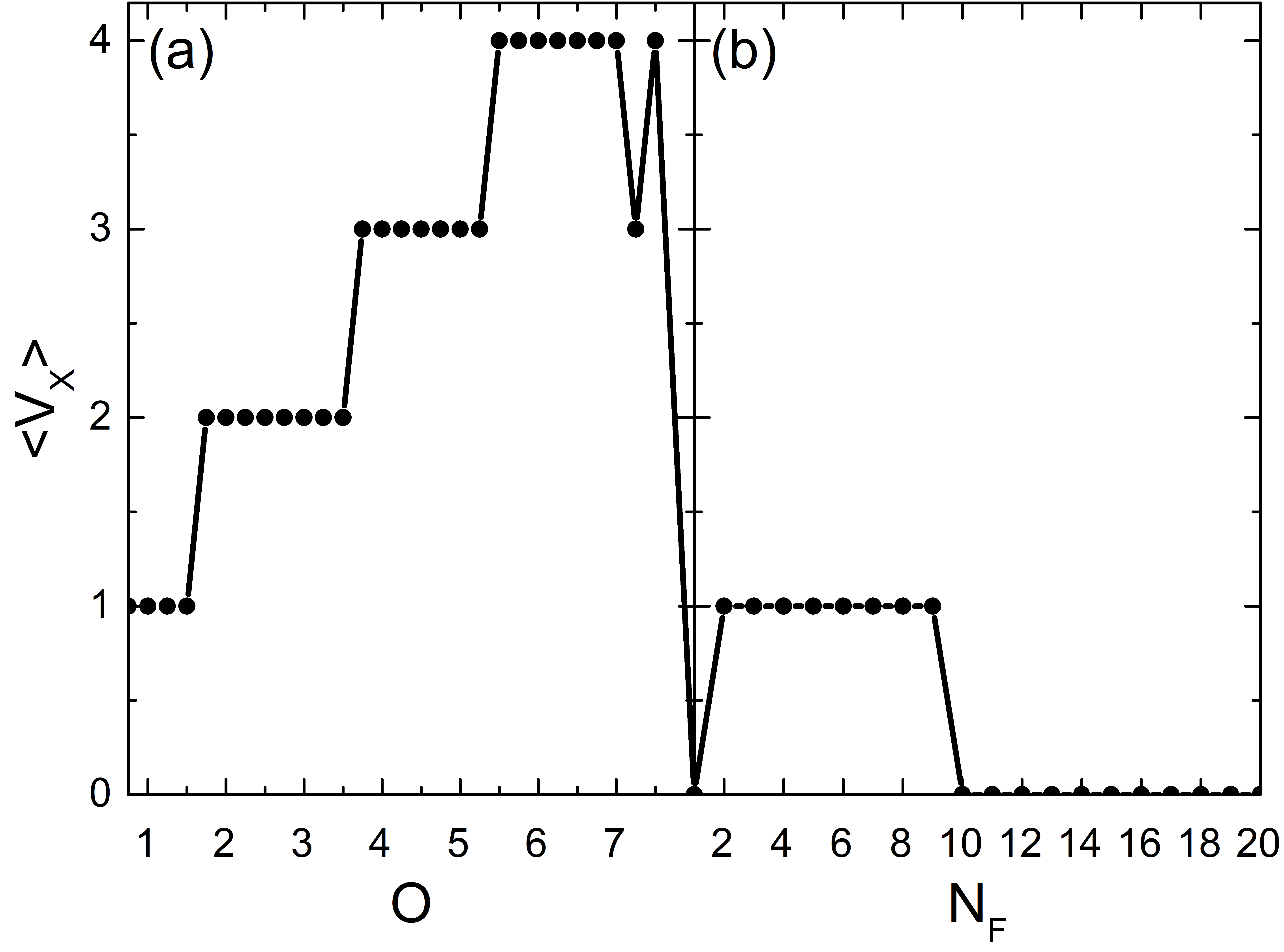}
    \caption{
$\langle V_x\rangle$ vs (a) $O$ for $N_F=4$ and (b) $N_F$ for $O=1.0$
in samples with biharmonic driving,      
$\alpha_m/\alpha_d = 0.5$, and
$A=B=2.0$.
Due to the normalization, $\langle V_x\rangle$ provides a measure of the
number of funnel plaquettes spanned by the translating trajectory during each
ac drive cycle.
}
    \label{Fig14}
\end{figure}
    
When its value is quantized,
$\langle V_x\rangle$ provides a measurement of the number of funnel
plaquettes spanned by a translating trajectory during each ac drive cycle.
We can use this measure to
explore the effects of varying the funnel parameters
out to their extreme
values.
In Fig.~\ref{Fig14}(a) we plot
$\left\langle V_x\right\rangle$
versus $O$
under biharmonic driving with $A=B=2.0$ at $N_F=4$, where we find
a clear increase in the skyrmion velocity as the funnel opening becomes larger.
The ratchet effect increases with increasing $O$
due to the increased probability for the skyrmion to reach
the next funnel along the easy direction.
When $O$ reaches its maximum size of $O=8.0$, however,
the funnel structure vanishes and thus the skyrmion ratchet
effect also disappears.
Abrupt vanishing of a ratchet effect with funnel aperture is common in
funnel ratchet systems.
When we fix $O=1.0$ and vary $N_F$,
Fig.~\ref{Fig14}(b)
shows that the number of funnels visited decreases with increasing $N_F$
since the funnel walls become steeper as each individual funnel becomes
shorter.
When $N_F>10$, the
skyrmion ratchet is lost for this choice of driving amplitude, $A=B=2.0$.

\begin{figure}[h]
    \centering
    \includegraphics[width=1.0\columnwidth]{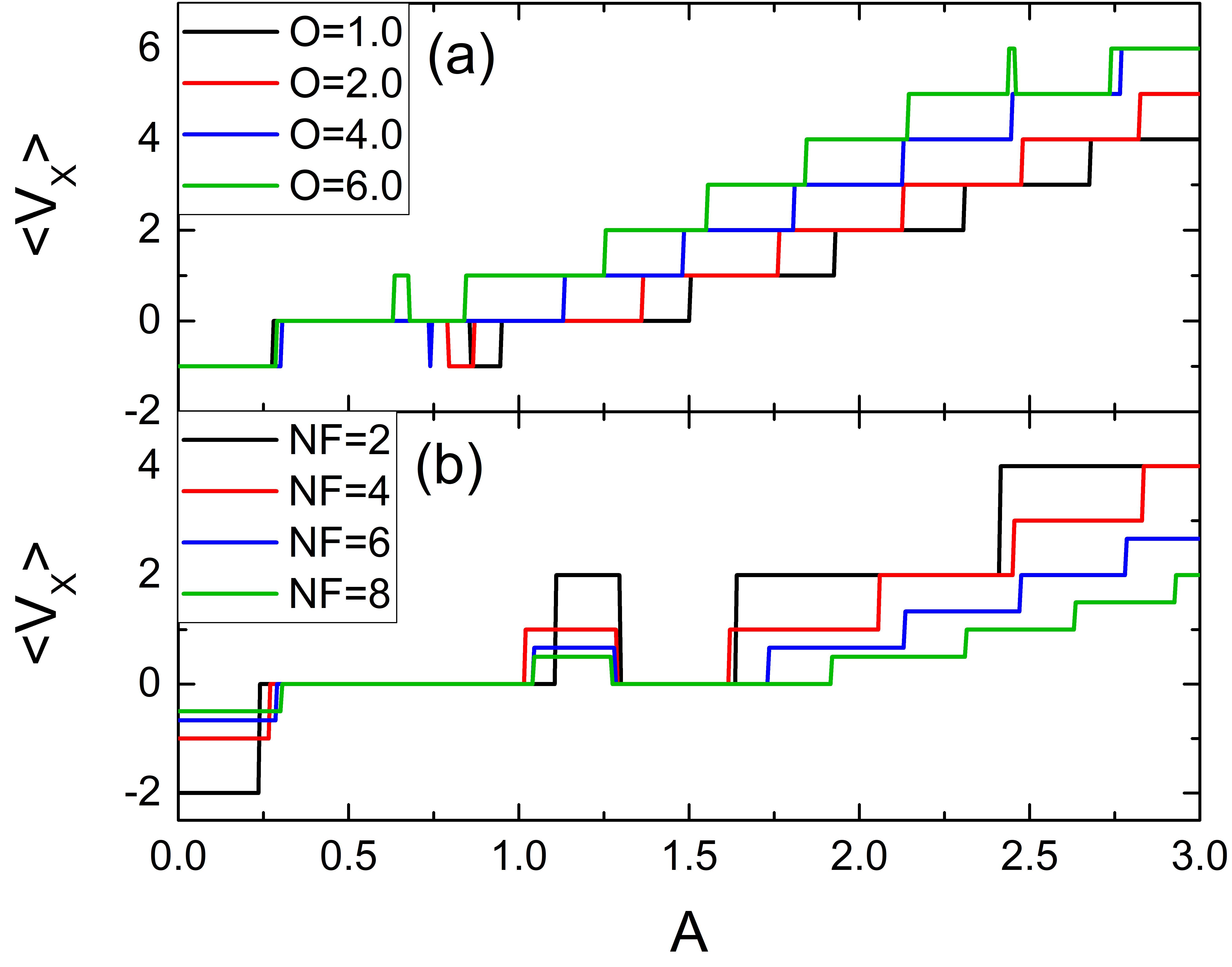}
\caption{
$\langle V_x\rangle$ vs $A$ in samples with biharmonic driving
and fixed $B=1.5$ at $\alpha_m/\alpha_d=0.5$.
(a) Varied $O$ for fixed $N_F=4$.
(b) Varied $N_F$ for fixed $O=0.28$.
To facilitate comparison, all velocities are normalized using $S=5$ even
though $S$ varies when $N_F$ is changed.
    }
    \label{Fig15}
\end{figure}

\begin{figure}[h]
   \centering
   \includegraphics[width=1.0\columnwidth]{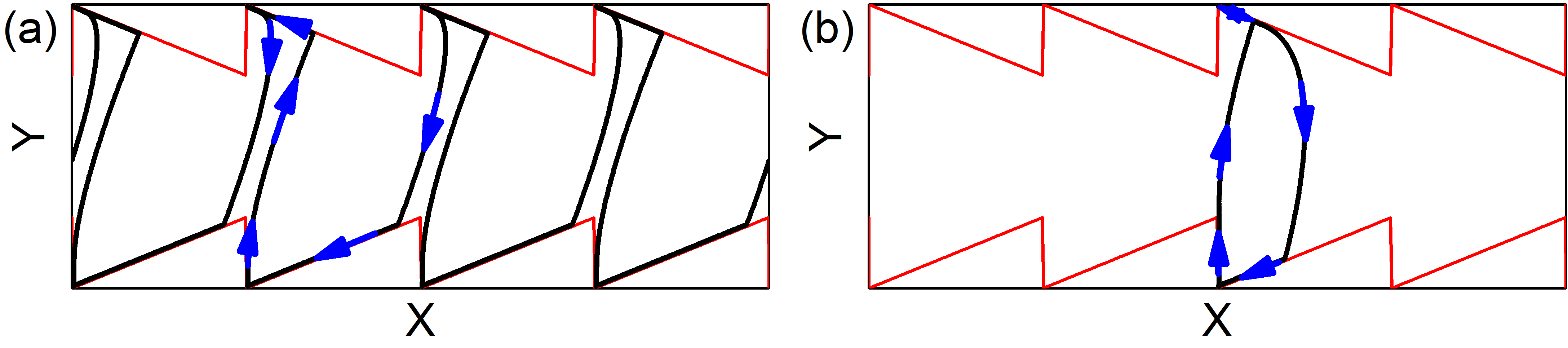}
   \includegraphics[width=0.5\columnwidth]{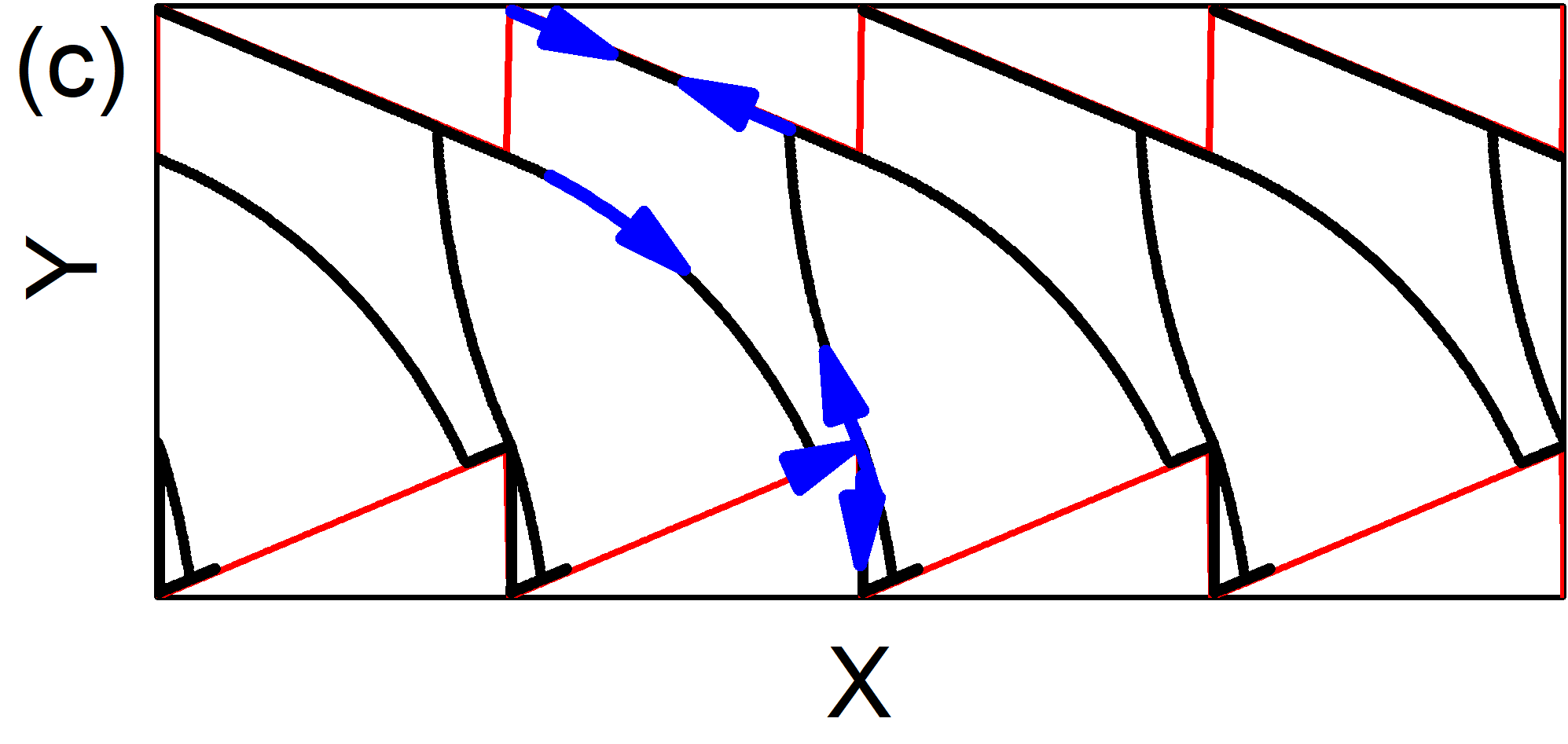}
\caption{Funnel wall (red lines) and the skyrmion trajectory (black lines)
in a system with  
$\alpha_m/\alpha_d = 0.5$ and a biharmonic ac drive
with $B=1.5$, $N_F=4$, and $O=4.0$. 
(a) At $A=0.25$, the skyrmion translates in the $-x$ direction due to
the Magnus-induced transverse ratchet effect.
(b)
At $A=0.5$, there is an intermediate state with no dc motion.
(c)
At $A=1.2$, the skyrmion translates in the $+x$ direction due to the
parallel ratchet effect.
}
        \label{Fig16}
    \end{figure}

In Fig.~\ref{Fig15} we plot $\langle V_x\rangle$ versus $A$
for a biharmonic ac drive system with fixed $B=1.5$ at $\alpha_m/\alpha_d=0.5$. 
An ac drive of this type
causes a free skyrmion to execute an asymmetric orbit,
and this asymmetry enhances the ratchet effect.
In the funnel geometry, the resulting skyrmion behavior
consists of a combination of the 
effects
found for unidirectional ac drives.
For very low values of $A$, the skyrmion
moves in the $-x$ direction due to the strong
Magnus-induced transverse ratchet
which appears at $B=1.5$, as illustrated in Fig.~\ref{Fig16}(a) at $A=0.25$.
For higher values of $A$, the parallel ratchet mechanism becomes
dominant and the
skyrmion moves in the $+x$ direction,
as shown in Fig.~\ref{Fig16}(c) at $A=1.2$.
An intermediate state appears between these regimes
when the Magnus-induced transverse ratchet and the 
parallel ratchet cancel each other, producing a localized
orbit with no dc motion, as shown in Fig.~\ref{Fig16}(b) at $A=0.5$.

\begin{figure}
  \centering
  \includegraphics[width=1.0\columnwidth]{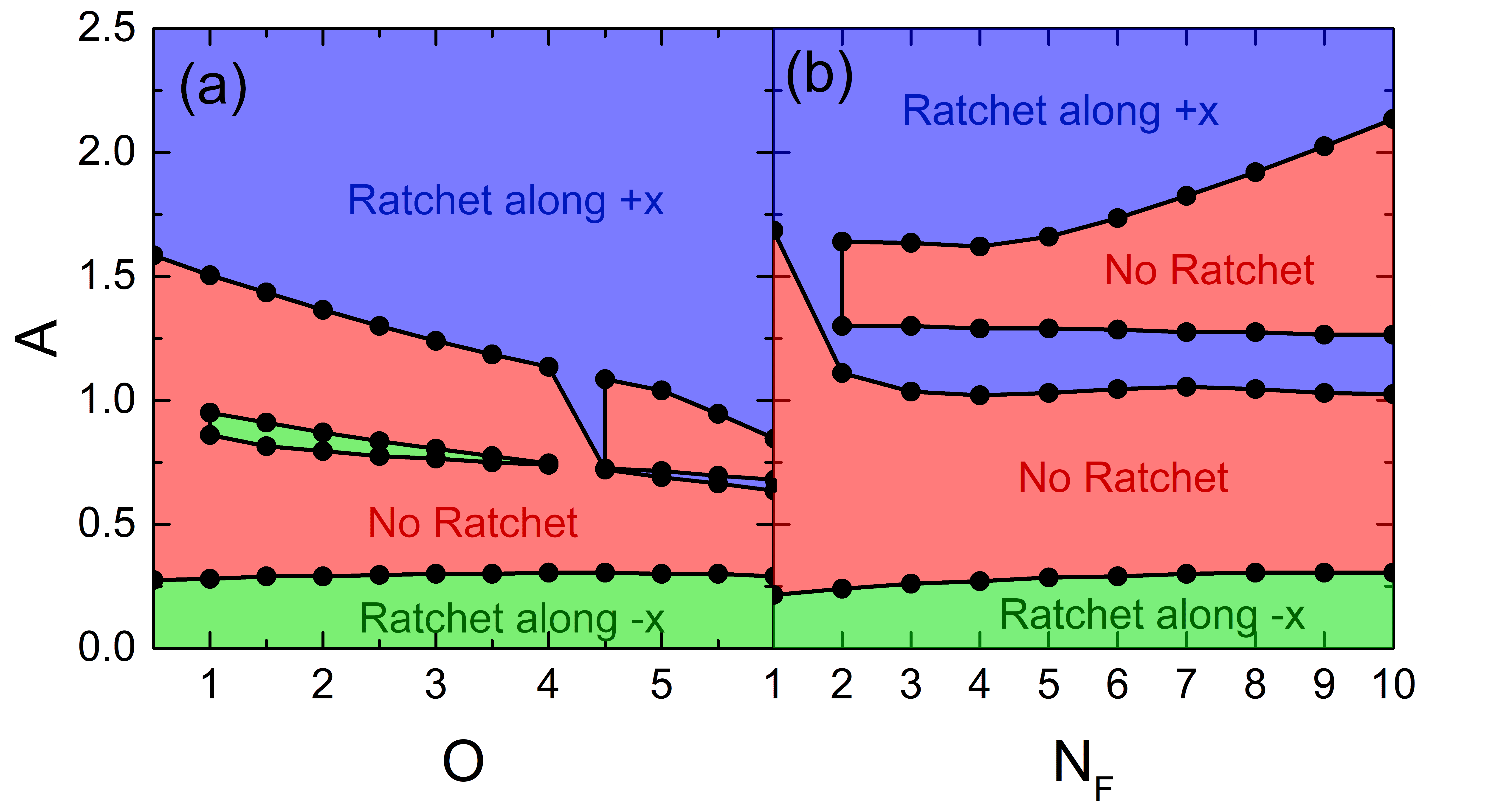}
  \caption{Dynamic phase diagrams for samples with
    biharmonic driving, $B=1.5$, and
    $\alpha_m/\alpha_d=0.5$ 
    as a function of (a) $A$ vs $O$ with $N_F=4$ and (b) $A$ vs $N_F$
    with $O=0.28$.
    Green areas indicate ratcheting motion in the hard or $-x$ direction,
    purple areas have ratcheting motion in the easy or $+x$ direction,
    and red areas show no ratchet effect.
  }
  \label{Fig17}
\end{figure}

In Fig.~\ref{Fig17}(a) we plot a dynamic phase diagram as a function
of $A$ versus $O$ for a
biharmonic system with $B=1.5$, $\alpha_m/\alpha_d=0.5$,
and $N_F=4$, while in Fig.~\ref{Fig17}(b) we show a similar dynamic
phase diagram as a function of $A$ versus $N_F$ for the same system at fixed
$O=0.28$. We find a reversal of the ratchet direction from $-x$
to $+x$
as $A$ increases, along with
multiple 
reentrant pinning regions in which ratcheting
behavior does not occur.
Figure~\ref{Fig17}(a) shows a window of $-x$ direction ratcheting
over the range $1.0<O<4.0$ in the middle of the pinned region when
$A \approx 0.8$. A similar window of $+x$ direction ratcheting appears
in the same region of $A$ for $O>4.5$. In Fig.~\ref{Fig17}(b),
we find a pinned phase for $N_F\geq 2$ close to $A=1.3$ where the
$-x$ ratchet motion ceases, followed by the emergence of ratcheting
motion in the $+x$ direction for higher $A$. Judicious selection of both
$O$ and $N_F$ determines the number and type of ratcheting phases that can
be accessed as $A$ is varied.

\section{Guided motion using AC drives}

\begin{figure}[h]
    \centering
    \includegraphics[width=1.0\columnwidth]{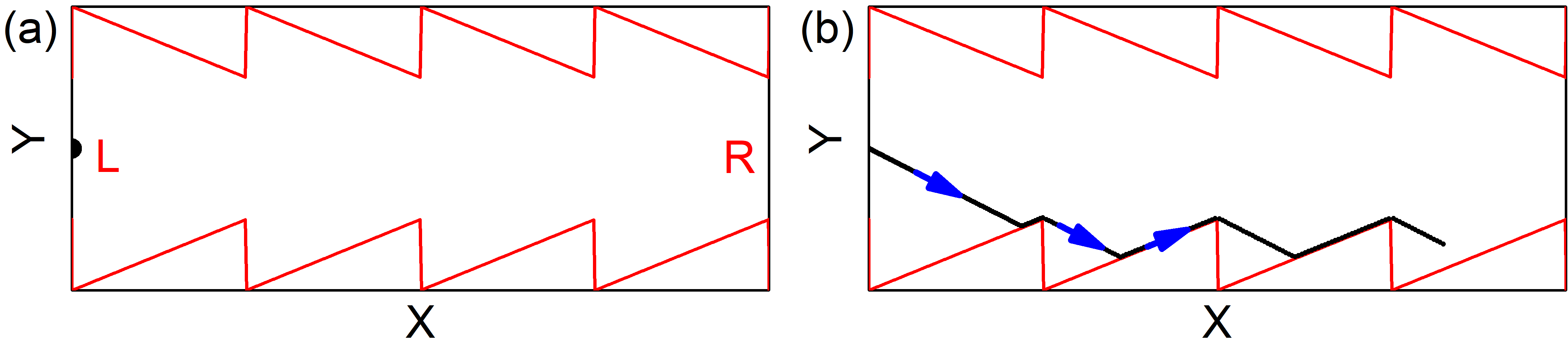}
    \includegraphics[width=1.0\columnwidth]{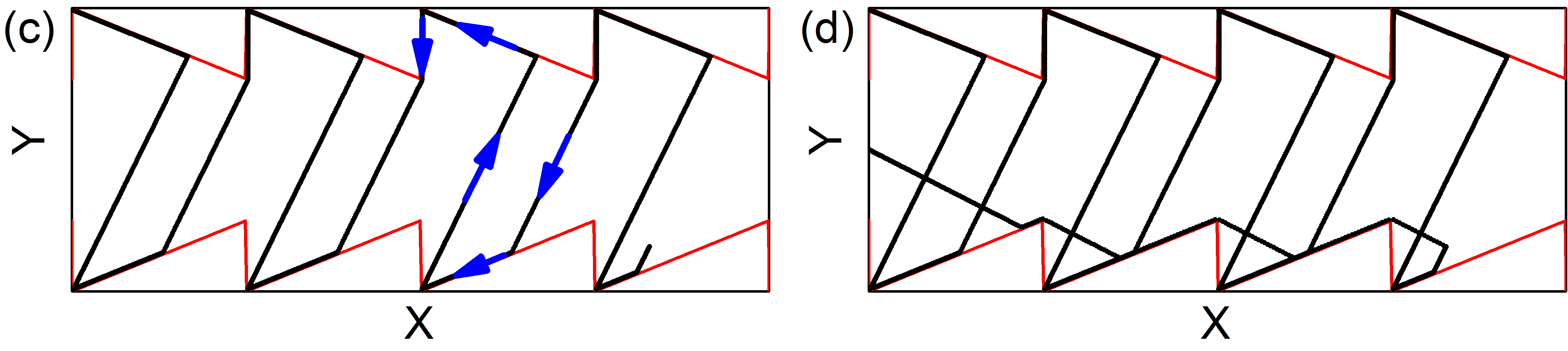}
    \caption{Funnel wall (red lines) and the skyrmion trajectory (black lines)
      under biharmonic driving
in a system with $N_F=4$, $O=4.0$, and $\alpha_m/\alpha_d=0.5$ illustrating
trajectory guiding from the left funnel, marked L, to the right funnel,
marked R, and back.
(a) Skyrmion starting position in funnel L at $t=0$.
(b) First stage of operation with $A=2.0$ and $B=0.0$,
where the skyrmion moves from $L$ to $R$.
(c) Second stage of operation with $A=0.0$ and $B=2.0$,
where the skyrmion moves from $R$ back to $L$.
(d) The complete trajectory.}
    \label{Fig18}
\end{figure}

\begin{figure}[h]
    \centering
    \includegraphics[width=1.0\columnwidth]{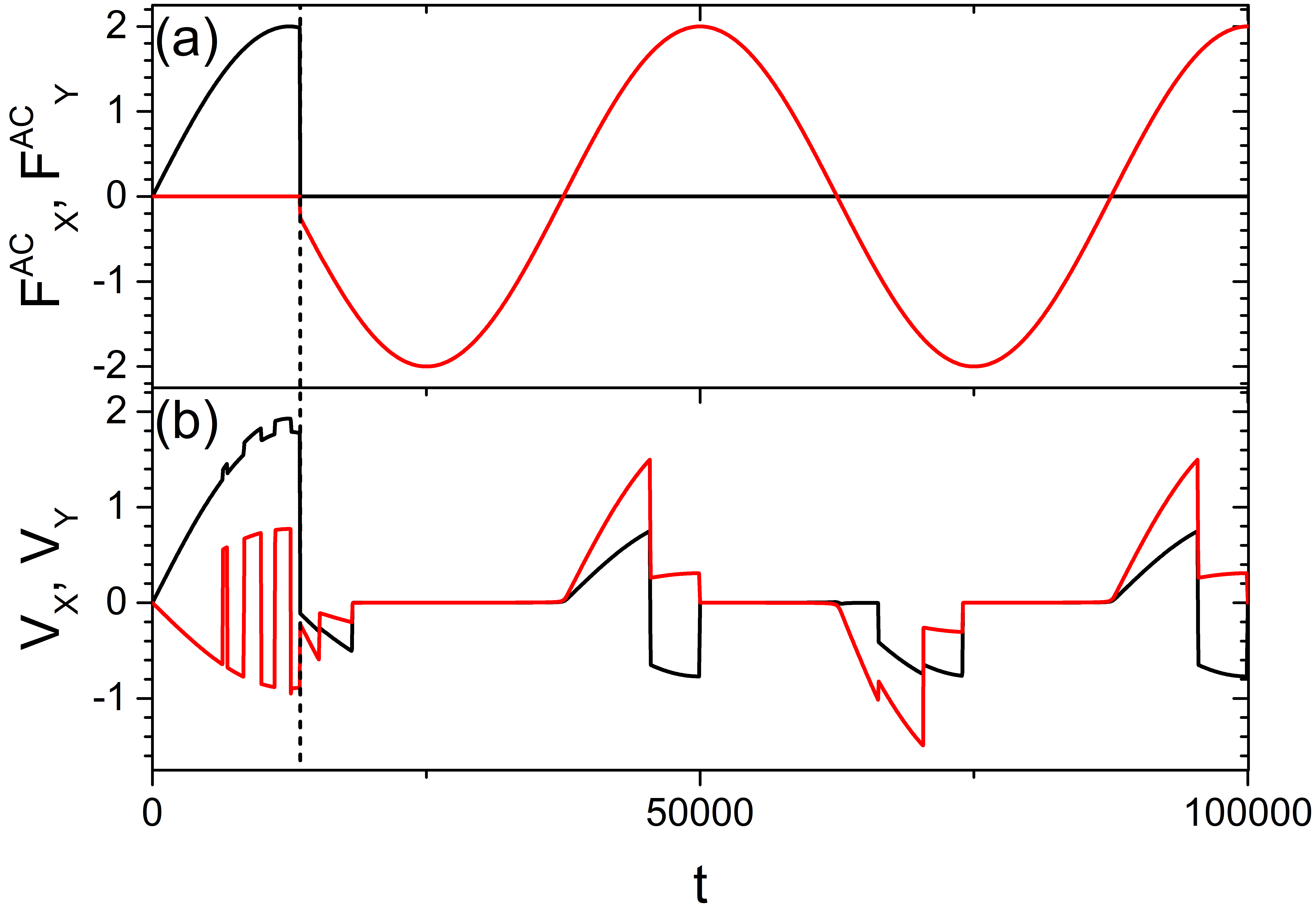}
\caption{
(a) The applied ac drive forces, $F^{AC}_x$ (black) and  $F^{AC}_y$ (red),
versus time $t$
for the guided skyrmion motion in Fig.~\ref{Fig17}.
The time axis is truncated at $t=100,000$, but the same ac driving is applied
at later times to achieve the transport in Fig.~\ref{Fig17}.
(b) The instantaneous skyrmion velocities
$V_x$ (black) and $V_y$ (red) versus $t$ under the driving shown in panel (a).
The vertical dashed line indicates the time at which the ac drive
direction is changed.
}
    
    \label{Fig19}
\end{figure}

In the previous section,
we showed that a skyrmion can be set into motion in the easy ($+x$) or
hard ($-x$) direction of the funnel axis depending on the orientation
of the ac driving force.
We now consider a situation in which
a skyrmion must be guided from an initial funnel
to another funnel and must then return to its 
original position under only ac driving.
This process is
relevant to spintronics devices
in which a skyrmion
acting as an information carrier
is guided through the sample
in order to transmit information.
In Fig.~\ref{Fig18}(a) we
illustrate the sample in which the skyrmion will move from
the left funnel labeled L to the right funnel labeled R and back again.
In Fig.~\ref{Fig19}(a) we plot the
applied ac drive 
signals $F^{AC}_x$ and $F^{AC}_y$
used to produce this motion as a function of time,
and in Fig.~\ref{Fig19}(b) we show
the corresponding instantaneous skyrmion velocities $V_x$
and $V_y$.
In order to guide the skyrmion along the easy
direction towards funnel $R$,
we apply ac driving in the $x$ direction with $A=2$ and $B=0$ for 
$13500$ time steps.
The skyrmion trajectory for this interval
is illustrated in Fig.~\ref{Fig18}(b).
As shown in
Fig.~\ref{Fig19},
only $1/4$ of an ac drive cycle is sufficient to bring the skyrmion
to funnel $R$ due to the high efficiency of motion in the easy direction.
During this time interval, Fig.~\ref{Fig19}(b) indicates
that the skyrmion velocity 
$V_x$ is always positive.
In order to 
guide the skyrmion back to funnel $L$,
we switch the ac driving into the $y$ direction
with $A=0$ and $B=2$ in order to
produce a Magnus-induced transverse ratchet
effect that translates
the skyrmion in the $-x$ direction.
The skyrmion trajectory for this interval
of time, which extends to $t=200000$,
is illustrated in Fig.~\ref{Fig18}(c).
The driving in this second stage of motion must be applied for a much
longer time interval since the motion in the hard direction is relatively
inefficient compared to motion in the easy direction.
This is highlighted in Fig.~\ref{Fig19}(b) where the skyrmion velocity $V_x$
drops to zero multiple times during the motion in the hard direction back
towards funnel L. We even observe time intervals in which the skyrmion moves
in the $+x$ direction, away from funnel L, before reversing direction and
moving back toward funnel L.
The complete skyrmion 
trajectory is shown in Fig.~\ref{Fig18}(d).
We expect that similar results could be achieved using 
asymmetric ac drives in which neither $A$ nor $B$ is zero.
As mentioned in Section V,
for a fixed $B$,
the skyrmion moves in the $-x$ direction
for low values of $A$ and in the $+x$ direction for
high values of $A$.
If asymmetric ac drives of this type were used to achieve
the type of transport illustrated in Fig.~\ref{Fig18},
the trajectories would be more 
complex.

\section{Discussion}

Using a combination of funnel geometries and
ac driving, it is possible to control the skyrmion direction of
motion accurately.
When the ac drive is applied parallel to
the funnel axis,
we find easy or $+x$ direction ratcheting,
while ratcheting in the hard or $-x$ direction
occurs for an ac drive applied perpendicular to the funnel 
axis.
The ratchet motion in the easy direction of the funnel
has been observed in a wide
variety of systems with overdamped particles 
\cite{wang_ratchets_2002,de_souza_silva_controlled_2006,libal_dynamics_2006,lee_observation_2005,villegas_experimental_2005}, 
but ratcheting motion in the hard direction
appears only for particles with strong Magnus 
terms \cite{reichhardt_magnus-induced_2015}.
This feature can be exploited to build devices 
in which the direction of the skyrmion transport
must be controlled precisely.
A future interesting aspect to explore
is the influence of temperature.
It is well-known that thermal 
effects can modify phase transition points and in some cases
can completely destroy them
\cite{reichhardt_thermal_2018}.
Thus, it is possible that the ratcheting motion along the hard direction
could be destroyed by strong 
thermal fluctuations,
since our results indicate that this motion is not very efficient.
Our simulation employs a particle-based skyrmion model
\cite{lin_particle_2013};
however, skyrmions also have 
internal structure that can be excited by applied ac
drives, or that can be deformed by an interaction with 
a wall barrier \cite{iwasaki_current-induced_2013}.
Such effects could be further 
explored in continuum-based simulations.
We considered the dynamics of a single skyrmion,
but we expect our results to be general for
the case of multiple skyrmions provided the skyrmion density 
is sufficiently low.
For multiple interacting skyrmions at higher densities, the skyrmion-skyrmion 
interactions cannot be neglected and could produce
significant modifications of the dynamics due to
collective 
effects \cite{reichhardt_nonequilibrium_2018,reichhardt_depinning_2016}.
Our results should be applicable not only to skyrmions, but to other
magnetic textures exhibiting significant Magnus effects, such as merons
\cite{Gobel21a}.

\section{Summary}
In this work we investigated the skyrmion behavior at zero temperature
in a funnel geometry under applied ac driving.
We show that for ac driving applied either parallel or perpendicular to
the funnel axis,
the skyrmion
can undergo net dc transport
along either the easy or the hard direction of the funnel.
For ac driving applied parallel to the funnel axis,
the skyrmion moves in the easy direction
with a quantized velocity
that increases monotonically with increasing ac drive amplitude.
In contrast, when the ac driving
is applied perpendicular to the funnel axis,
the skyrmion flows in the hard direction at a constant average velocity
due to a
Magnus-induced transverse ratchet effect.
Increasing the relative importance of the Magnus term compared to the
damping term
only weakly affects the easy direction ratcheting,
but can destroy the dc motion
of the transverse ratchet effect.
For biharmonic driving with equal drive amplitudes
both parallel
and perpendicular to the funnel axis,
a reentrant pinning phase appears that depends on the size of the funnel 
opening and the number of funnels.
Under asymmetric
biharmonic ac driving, where the perpendicular ac drive amplitude is held
fixed and the parallel ac drive amplitude $A$ is varied,
we find a combination of the effects observed for separate ac driving along
each axis.
For low values of $A$, the skyrmion moves along the hard direction
due to the prominent transverse ratchet 
effect; however,
for higher values of $A$,
a transition occurs to motion
along the easy direction
due to the strengthening of the parallel ratchet effect.
These results may be useful for future devices since 
they indicate that the skyrmion motion
in a funnel geometry can be controlled using only ac driving.
To illustrate this, we demonstrate that a skyrmion can be transported to
a series of predetermined locations by varying the ac driving over time,
simulating a data transfer process.

\acknowledgments
This work was supported by the US Department of Energy through
the Los Alamos National Laboratory.  Los Alamos National Laboratory is
operated by Triad National Security, LLC, for the National Nuclear Security
Administration of the U. S. Department of Energy (Contract No. 892333218NCA000001).
N.P.V. acknowledges
funding from
Funda\c{c}\~{a}o de Amparo \`{a} Pesquisa do Estado de S\~{a}o Paulo - FAPESP (Grant 2017/20976-3).

\bibliography{mybib}
\end{document}